\documentclass[10pt,twocolumn,superscriptaddress,longbibliography,preprintnumbers,floatfix,nofootinbib,aps,prd]{revtex4-2}

\usepackage{url}
\usepackage{xspace}
\usepackage{dsfont}
\usepackage{amssymb}
\usepackage{amsmath}
\usepackage{graphicx}
\usepackage[caption=false]{subfig}
\usepackage[colorlinks=true,citecolor=blue]{hyperref}
\usepackage{multirow}
\usepackage{hhline}
\usepackage{xcolor}
\usepackage{comment}
\usepackage{cleveref}
\crefname{appendix}{Appendix}{Appendices}
\usepackage[acronym]{glossaries}
\setkeys{glslink}{hyper=false}
\usepackage{booktabs}
\usepackage{makecell}
\usepackage{siunitx}
\usepackage[normalem]{ulem}
\DeclareSIUnit\barn{b}

\graphicspath{{Plots/}}

\newacronym{ggF}{ggF}{gluon--gluon fusion}
\newacronym{VBF}{VBF}{vector boson fusion}
\newacronym{VH}{VH}{VH}
\newacronym{ttH}{ttH}{ttH}
\newacronym{CATHODE}{CATHODE}{Classifying Anomalies THrough Outer Density Estimation}
\newacronym{BDT}{BDT}{Boosted Decision Tree}
\newacronym{SR}{SR}{Signal Region}
\newacronym{SB}{SB}{Sideband Region}
\newacronym{SM}{SM}{Standard Model}
\newacronym{BSM}{BSM}{Beyond the Standard Model}
\newacronym{SUSY}{SUSY}{Supersymmetry}
\newacronym{QCD}{QCD}{quantum chromodynamics}
\newacronym{VAE}{VAE}{Variational Autoencoder}
\newacronym{NF}{NF}{Normalizing Flow}
\newacronym{SIC}{SIC}{Significance Improvement Characteristic}
\newacronym{AUC}{AUC}{area under the curve}
\newacronym{ROC}{ROC}{Receiver Operating Characteristic}
\newacronym{TPR}{TPR}{True Positive Rate}
\newacronym{BCE}{BCE}{Binary Cross-Entropy}
\newacronym{ML}{ML}{machine learning}
\newacronym{MLP}{MLP}{Multilayer Perceptron}
\newacronym{KL}{KL}{Kullback-Leibler divergence}
\newacronym{ID}{ID}{In Dataset}
\newacronym{IP}{IP}{Interpolation}
\newacronym{EP}{EP}{Extrapolation}
\newacronym{MC}{MC}{Monte Carlo}
\newacronym{cwola}{CWoLa}{Classification Without Labels}
\begin{document}

\title{Signal-Aware Contrastive Latent Spaces for Anomaly Detection}

\author{Runze Li}%
\email{runze.li@yale.edu}
\affiliation{Department of Physics, Yale University, New Haven, CT 06511, USA}

\author{Benjamin Nachman}%
\email{nachman@stanford.edu}
\affiliation{Department of Particle Physics and Astrophysics, Stanford University, Stanford, CA 94305, USA}
\affiliation{Fundamental Physics Directorate, SLAC National Accelerator Laboratory, Menlo Park, CA 94025, USA}

\author{Dennis Noll}%
\email{nollde@stanford.edu}
\affiliation{Department of Particle Physics and Astrophysics, Stanford University, Stanford, CA 94305, USA}
\affiliation{Fundamental Physics Directorate, SLAC National Accelerator Laboratory, Menlo Park, CA 94025, USA}

\date{\today}
\begin{abstract}

  High-dimensional feature spaces in particle physics events pose a fundamental challenge to density-estimation-based weakly supervised anomaly detection, whose fidelity degrades rapidly with an increasing number of dimensions.
  We propose a signal-aware latent space construction using supervised contrastive learning trained on simulated Standard Model backgrounds and a diverse set of hypothesized Beyond the Standard Model (BSM) signals.
  The resulting latent space is low-dimensional, regularized, and signal-sensitive, enabling high-fidelity density estimation for downstream weakly supervised anomaly detection.
  We demonstrate the approach in a diphoton final state, testing sensitivity across a broad range of BSM scenarios including supersymmetry models, extended Higgs sectors, heavy neutral resonances, and flavor-changing neutral currents.
  For signals represented in the contrastive training data, the method can elevate discovery sensitivity from previously inaccessible levels to the discovery regime.
  Critically, the approach retains sensitivity to BSM models not present during training: interpolation and extrapolation to unseen signal topologies yield substantial improvements in expected significance compared to a background-only baseline.
  By bridging supervised latent space embedding with weakly supervised anomaly detection, this strategy offers a viable path toward anomaly detection in high-dimensional feature spaces at the LHC and beyond.

\end{abstract}

\maketitle

\section{Introduction}
\label{sec:introduction}

Anomaly detection is the automated search for new phenomena and is an essential set of tools to complement the existing search program at colliders and other experimental facilities~\cite{Kasieczka:2021xcg,Karagiorgi:2021ngt,Belis:2023mqs}.
One well-studied class of anomaly detection strategies is weakly supervised anomaly detection, which offers asymptotically optimal sensitivity to new physics~\cite{Metodiev:2017vrx,Collins:2018epr,Nachman:2020lpy}.
This approach has been applied to recorded collision data from both the ATLAS~\cite{ATLAS:2020iwa,ATLAS:2025obc} and CMS~\cite{CMS:2024nsz,cmscollaboration2025machinelearningtechniquesmodelindependentsearches,Gambhir:2025afb,bhimji2025omnilearnedfoundationmodelframework} experiments.

Weakly supervised anomaly detection approaches, such as \gls{CATHODE}~\cite{Hallin:2021wme}, consist of two parts: building a background-only reference through density estimation and comparing the reference to real data that may include a signal contribution.
Both the density estimation and the classification become more challenging when the dimensionality of the feature space is increased~\cite{finke_back_2023,Freytsis:2023cjr,buhmann2024phasespaceresonantanomaly}.
A number of solutions to this scaling challenge have been proposed.
For the classifier, shallow methods can be more resilient than deep learning, especially for tabular data~\cite{grinsztajn2022treebasedmodelsoutperformdeep}.
Pre-training has also proven effective, both for point-cloud representations of the data~\cite{Mikuni:2025tar} and for improving the precision of the background estimation~\cite{PhysRevD.111.L051504}.
Intuitively, these strategies use effectively lower-dimensional versions of the data through regularized function approximators.
An alternative strategy is to directly compress the data without restricting the function space.
This was explored, for example, with symmetry-based contrastive learning in Ref.~\cite{Dillon:2022tmm}.
While completely signal- and simulation-agnostic, the symmetry constraints may not yield a sufficiently expressive latent representation of the data.

We revisit contrastive learning to enable high-dimensional anomaly detection through compression.
To improve the learned representation, we employ supervised contrastive learning trained on simulated \gls{SM} backgrounds and a diverse set of hypothesized \gls{BSM} signals.
This allows for the construction of a latent representation that is explicitly sensitive to the differences between physics models in the feature space.

Previous studies considered a similar setup for constructing a learned latent space using physics processes in the contrastive training, but limited the processes to backgrounds~\cite{contrastive_embedding_bkgonly}.
The novelty of our work is the inclusion of a wide variety of signal processes in the contrastive training.
Our hypothesis is that the inclusion of signals will not only improve the sensitivity of the latent space for the signals used in training, but will also generalize to unseen signals through interpolation across parameter spaces and extrapolation to entirely new signal topologies.
Building on Ref.~\cite{haxadv1}, which demonstrated weakly supervised anomaly detection in a diphoton final state but was limited to $\sim10$ features by the fidelity of the density estimation, we show that our approach can both lift this dimensionality constraint and substantially improve the anomaly detection sensitivity.

This paper is organized as follows.
\Cref{sec:dataset} describes the simulated dataset and the signal models used for training and evaluation.
\Cref{sec:strategy} introduces the contrastive embedding and the \gls{CATHODE} anomaly detection pipeline.
\Cref{sec:results} presents the results, including the structure of the latent space, sculpting checks, and the anomaly detection sensitivity across different embedding configurations.
\Cref{sec:conclusion} summarizes the findings and discusses future directions.

\section{Dataset}
\label{sec:dataset}

The analysis uses the same simulated events as Ref.~\cite{haxadv1}, corresponding to proton--proton collisions at a center-of-mass energy of $\sqrt{s} = \SI{13}{\TeV}$ with an integrated luminosity of $\mathcal{L} = \SI{137}{\per\femto\barn}$.
Event generation is performed with \textsc{MadGraph5\_aMC@NLO}~v3.5.9~\cite{Alwall:2011uj} and \textsc{Pythia}~v8.312~\cite{Sjostrand:2006za,Sjostrand:2014zea} at leading order in \gls{QCD}, with detector simulation via \textsc{Delphes}~v3.5.0~\cite{deFavereau:2013fsa,Mertens:2015kba} using a simplified ATLAS detector layout.
Full details of the generation configuration are given in Ref.~\cite{haxadv1}.

The event pre-selection broadly follows the standard selection of $H \rightarrow \gamma\gamma$ analyses.
The diphoton candidate is formed from the pair of the three highest transverse momentum ($p_T$) photons whose invariant mass is closest to the Higgs boson mass ($m_H = \SI{125}{\GeV}$).
Events are required to satisfy a diphoton trigger, requiring the leading and sub-leading photons of the diphoton candidate to have $p_T > \SI{35}{\GeV}$ and $p_T > \SI{25}{\GeV}$, respectively, or alternatively a single-photon trigger with $p_T > \SI{140}{\GeV}$.
To enhance the selection of photons from a resonant decay, the leading and sub-leading photons from the diphoton candidate must satisfy $p_T / m_{\gamma\gamma} > 0.4$ and $p_T / m_{\gamma\gamma} > 0.3$, respectively.
The diphoton invariant mass is required to lie within $\SI{105}{\GeV} < m_{\gamma\gamma} < \SI{160}{\GeV}$.
Jets are reconstructed using the anti-$k_T$ algorithm~\cite{Cacciari:2005hq, Cacciari:2011ma, Cacciari:2008gp} with $R = 0.4$.
Each event is represented by a set of reconstructed physics objects as described in \cref{sec:embedding}.

The \gls{SM} background comprises non-resonant diphoton continuum processes ($\gamma\gamma + \text{jets}/\ell\ell/t\bar{t}$) and resonant \gls{SM} Higgs production via \gls{ggF}, \gls{VBF}, associated production with a weak boson ($VH$), and production with a top-quark--antiquark pair ($t\bar{t}H$), with $H \rightarrow \gamma\gamma$.

Compared to Ref.~\cite{haxadv1}, the set of \gls{BSM} signal processes is substantially expanded for the signal-aware supervised contrastive training and to benchmark the sensitivity across a broad range of topologies.
The full set of signal models is summarized in \cref{tab:signals}.
Only $H \rightarrow \gamma\gamma$ decays are realized for the Higgs boson in all signal processes.

The simulated events are randomly split into two sets.
The first set is used to train the encoder model, and the second set is used to construct pseudo data corresponding to the target integrated luminosity.
To construct the pseudo data, the \gls{SM} background events are randomly sampled according to weights derived from their production cross section and generator-level event weights such that the pseudo data has the expected number of events.
Different \gls{BSM} signal events can then be injected into the pseudo data to test the sensitivity of our method.
In an analysis of recorded data, the first dataset would correspond to the \gls{MC} simulated events, and the second dataset would be the recorded data from the experiment.

\begin{table*}[htbp]
  \centering
  \caption{Signal and background processes used in this study. Mass parameters are given in GeV. RPV and RPC denote R-parity violating and R-parity conserving scenarios, respectively. The chargino-neutralino model features prompt decays $\tilde{\chi}_1^\pm \rightarrow W^\pm \tilde{\chi}_1^0$ and $\tilde{\chi}_2^0 \rightarrow H \tilde{\chi}_1^0$. In the extended Higgs sector model, a heavy scalar $X$ is produced via \gls{ggF} and decays as $X \rightarrow S(\ell\ell)\, H$, where $S$ is a lighter \gls{BSM} scalar.
  \label{tab:signals}}

  \begin{tabular}{lccc}
    \toprule
    Process & Label & Group ID & Reference \\
    \midrule

    \acrshort{SM} non-resonant backgrounds & $\gamma\gamma + jjj / \ell\ell /  t\bar{t}$        & 0 & --- \\
    \midrule

    \multirow{4}{*}{\acrshort{SM} Higgs production}
    & \acrshort{ggF} & 1 & --- \\
    & \acrshort{VBF} & 2 & --- \\
    & \acrshort{VH} & 3 & --- \\
    & \acrshort{ttH} & 4 & --- \\
    \midrule

    \acrshort{SUSY} chargino-neutralino & \makecell{$(\chi_1^{\pm}\chi_2^{0})_m$\\ $m \in\{150,200,300,600\}$ } & -1 & \cite{collaboration_search_2020} \\
    \midrule

    Extended Higgs sector & $X \to S(\ell\ell)H$ & -2 & \cite{xsh_2024} \\
    \midrule

    \acrshort{SUSY} RPV stop & \makecell{$(\tilde t\,\tilde t)_m$ (RPV)\\ $m \in \{150,300,450\}$} & -3 & \cite{Monteux_2016} \\
    \midrule

    \acrshort{SUSY} RPV neutralino & \makecell{$(\chi_1^{\pm}\chi_1^{0})_m$ (RPV)\\ $m \in \{200,400,600\}$} & -4 & \cite{Monteux_2016} \\
    \midrule

    \acrshort{SUSY} RPC sbottom & \makecell{$(\tilde b\,\tilde b)_m$\\ $m \in \{500,1000,1200\}$}
    & -5 & \cite{SUSY-2018-31} \\
    \midrule

    Heavy neutral resonance & \makecell{$Y_m \to H\gamma$\\ $m \in \{200,300,400,500\}$} & -6 & \cite{Aad_2020} \\
    \midrule

    \multirow{2}{*}{FCNC}
    & $tH$ (FCNC)       & \multirow{2}{*}{-7} & \multirow{2}{*}{\cite{CMS_FCNC_2018}} \\
    & $t\bar t$ (FCNC)   &                    & \\
    \midrule

    \acrshort{SUSY} RPC stop & $\tilde t\,\tilde t$ & -8 & \cite{ATLAS_RPC_STOP_2020} \\

    \bottomrule
  \end{tabular}
\end{table*}

\section{Strategy}
\label{sec:strategy}

The method follows a two-stage approach.
In the first stage, a supervised contrastive encoder maps the high-dimensional event representation into a low-dimensional latent space that is regularized for density estimation while still being sensitive to \gls{BSM} signals.
In the second stage, the \gls{CATHODE} method~\cite{Hallin:2021wme} is applied in this latent space to perform weakly supervised anomaly detection through data-driven background estimation and classification.
The two stages are described in the following sections.

\subsection{Embedding}
\label{sec:embedding}

The embedding stage employs supervised contrastive learning~\cite{supervisedcontrastivelearning} to construct a low-dimensional latent space from the high-dimensional event representation.
An encoder is trained to pull events from the same physics processes together in the latent space while pushing events from different physics processes apart.
The latent space is designed to satisfy two requirements: sensitivity to \gls{BSM} signals, achieved by including a diverse set of signal models alongside \gls{SM} backgrounds in the contrastive training, and modelability by the downstream generative model, enforced through Kullback--Leibler regularization toward a unit Gaussian prior for each physics process.
The encoder architecture, loss function, and training procedure are described below; the resulting latent space structure is examined in \cref{sec:latent_space}.

The encoder is based on a particle transformer architecture similar to Ref.~\cite{qu2024particletransformerjettagging}, comprising four particle attention blocks and two class attention blocks, totaling approximately 1.1M trainable parameters.
For each event, 11 objects are used: four small-radius ($R=0.4$) jets, two large-radius ($R=1.0$) jets, two electrons, two muons, and $\mathrm{E}^{\mathrm{miss}}_\mathrm{T}$.
The overlap between the small-radius jets and large-radius jets is not removed.
Each object is described by 12 features, including its four-momentum, $b$-tagging score, two $n$-subjettiness variables $\tau_{32}$ and $\tau_{43}$ \cite{nsubjettiness}, and a five-component one-hot encoding indicating whether the object is a small-radius jet, large-radius jet, electron, muon, or $\mathrm{E}^{\mathrm{miss}}_\mathrm{T}$.
Padding with zeros is applied for missing objects and features ($n$-subjettiness for leptons, for example).
The resulting $11 \times 12$ feature tensor is processed by the particle transformer, and the output is concatenated with a high-level feature tensor encoding $\Delta R_{\gamma\gamma}$ and $p_{\mathrm{T}\gamma\gamma}$ through a \gls{MLP}.
The concatenated representation is then passed to the output block. 

Kinematic inputs are normalized before being fed into the model.
Specifically, transverse momenta and masses are transformed to a logarithmic scale, and each feature is then standardized by subtracting its mean and dividing by its standard deviation.
Normalization statistics for object-level features are computed over all objects ($11 \times N$), while event-level features are normalized individually ($1 \times N$).
This ensures that each feature type is normalized according to its relevant context and with respect to missing objects.

By construction, the photon four-momenta are excluded from the transformer input, so the learned latent space will not be directly correlated with $m_{\gamma\gamma}$.
This decorrelation simplifies the density estimation and interpolation task in the \gls{CATHODE} step and suppresses background sculpting that could arise from imperfect modeling of $m_{\gamma\gamma}$-dependent features~\cite{Hallin:2021wme}.
The trade-off is a potential loss of sensitivity, since these features cannot be used by the encoder.
Dedicated decorrelation techniques could be employed to retain more of the photon information while enforcing independence of $m_{\gamma\gamma}$ and the latent space.

The output block of the encoder is inspired by the \gls{VAE}~\cite{kingma2014autoencoding}: it predicts a mean and variance from which the latent vector is sampled via the reparameterization trick, encouraging a Gaussian latent space that is well suited for downstream density estimation and interpolation.
Similar approaches are used in Refs.~\cite{Hallin:2022eoq,haxadv1}.
During training, a two-layer \gls{MLP} projection head is applied after the output block; it is discarded at inference time.
This practice has been shown to improve the quality of the learned embedding~\cite{simclr,contrastive_embedding_bkgonly}.

The encoder is trained with a loss function that combines supervised contrastive learning with \gls{KL} regularization.
Given $N$ events in a batch $I$ with indices $i$ and physics-process labels $y_i$, let $z_i$ denote the latent vector from the encoder output block and $Z_i$ the corresponding projection-head output.
Define $A(i) \equiv I \setminus \{i\}$ and $P(i) \equiv \{\, p \in A(i) : y_p = y_i \,\}$ which is the set of all other events in the batch sharing the same label.
Unlike Ref.~\cite{supervisedcontrastivelearning}, which augments each event to create multiple views, we use each event only once and do not perform augmentations, as the training dataset is sufficiently large.
The total loss is
\begin{equation}
  \label{eq:loss}
  \begin{gathered}
  \mathcal{L}
    = \mathcal{L}_{\mathrm{con}} + \lambda\, \mathrm{KL}\!\left(z \,\middle\|\, \mathcal{N}(0,1)\right) \quad \text{with}\\
    \mathcal{L}_{\mathrm{con}}
    = \sum_{i\in I} \frac{-1}{|P(i)|} \sum_{p\in P(i)}
    \log \frac{\exp\!\left(Z_i \cdot Z_p / \tau\right)}
    {\sum_{a\in A(i)} \exp\!\left(Z_i \cdot Z_a / \tau\right)}
  \end{gathered}
\end{equation}
The first term ($\mathcal{L}_{\mathrm{con}}$) is the supervised contrastive loss of Ref.~\cite{supervisedcontrastivelearning}, which pulls events with the same label closer together in the latent space while pushing events with different labels apart.
The loss temperature $\tau$ is a hyperparameter which controls the regularization strength of the contrastive loss.
The second term is the \gls{KL} divergence between the distribution of the latent vector $z$, sampled from the Gaussian parameterized by the encoder, and a unit Gaussian prior.
This term acts as a regularizer and the hyperparameter $\lambda$ controls the regularization strength: larger values of $\lambda$ yield a more Gaussian latent space that is easier to model in the downstream \gls{CATHODE} step, at the cost of reduced influence from the contrastive loss.

From the encoder training set described in \cref{sec:dataset}, 100,000 events from each of the processes listed in \cref{tab:signals} are further split 60-20-20 into training, validation, and test sets.
For processes with multiple mass parameters, 100,000 events are drawn for each mass point.
Events are labeled according to their physics-process group: non-resonant $\gamma\gamma + X$, \gls{ggF}, \gls{VBF}, \gls{VH}, \gls{ttH}, and eight \gls{BSM} signal groups.
For \gls{BSM} processes, all mass points of a given model share the same label.
Four training configurations are explored to probe the generalization of the learned latent space:
\begin{itemize}
  \item \textbf{\gls{ID}}: all process groups are included in training;
  \item \textbf{\gls{IP}}: one mass point of a \gls{BSM} model is excluded;
  \item \textbf{\gls{EP}}: all mass points of a \gls{BSM} model are excluded;
  \item \textbf{Background only}: only \gls{SM} processes are included.
\end{itemize}
The \gls{ID} configuration, in which the encoder has access to all processes including those used for testing, is expected to yield the best performance.
The \gls{IP} configuration tests interpolation to unseen mass points within a known signal topology.
The \gls{EP} configuration evaluates extrapolation to entirely unseen signal models.
The background-only configuration, similar to Ref.~\cite{contrastive_embedding_bkgonly}, serves as a benchmark without direct signal information.

Training uses a batch size of 8192 with a sampler that ensures uniform representation of each process group per batch.
The Adam optimizer~\cite{adam} is employed with an initial learning rate of $2\times10^{-4}$, five epochs of linear warm-up, and cosine annealing.
Training runs for up to 100 epochs with early stopping based on a patience of 10 epochs.
The contrastive loss temperature is set to $\tau = 0.1$.
As the number of process groups increases, the latent space tends to be less regular due to the larger amount of information it must encode.
In order to have a smooth latent space distribution, the \gls{KL} weight $\lambda$ is optimized separately for the background-only configuration ($\lambda = 0.01$) and the IP and EP configurations ($\lambda = 0.1$).
In a practical application, this optimization is legitimate, since the contrastive encoder is trained on simulated event samples.
The latent space dimensionality is set to six.
While higher dimensions can improve encoder performance, they have been shown to make the downstream density estimation in the \gls{CATHODE} step more difficult.
A six-dimensional latent space with the optimized $\lambda$ regularization is found to be well modeled by the \gls{NF} while maintaining high sensitivity to different tested \gls{BSM} processes.

\subsection{CATHODE}
\label{sec:cathode}

To detect anomalies, the \gls{CATHODE} method~\cite{Hallin:2021wme}, which has been shown to be asymptotically optimal for this task, is employed.
As described in \cref{sec:dataset}, one half of the simulated events is used to train the encoder, while the other half is used to construct pseudo data for evaluating the sensitivity.
The \gls{CATHODE} method is applied to the latent representation of the pseudo data in a multi-step procedure.

First, events of the pseudo data are divided according to $m_{\gamma\gamma}$ into a \gls{SR}, where a potential signal contribution is expected, and a \gls{SB}, where no potential signal contribution is expected.
Then, an \gls{NF}, conditioned on $m_{\gamma\gamma}$, is trained with events from the \gls{SB} to learn the background density in the latent space.
The trained model is then used to generate synthetic background samples in the signal region by interpolating the learned density to $m_{\gamma\gamma}$ values from the \gls{SR}.
The regularized, near-Gaussian structure of the contrastive latent space facilitates high-fidelity density estimation, while its signal-sensitive construction ensures that genuine \gls{BSM} contributions remain distinguishable from the background estimate (see \cref{sec:latent_space}).
Second, a classifier is trained using the \gls{cwola} method~\cite{Metodiev:2017vrx} to distinguish the events from data and the background estimate in the signal region.
Events identified as data-like by this classifier are anomaly candidates which could then be leveraged with methods like bump hunting for a physics measurement (see e.g. Ref.~\cite{haxadv1}).

A critical requirement of the \gls{CATHODE} method is that the background estimate does not sculpt artificial bumps in the $m_{\gamma\gamma}$ spectrum, which could lead to false discoveries~\cite{Hallin:2021wme}.
The absence of such sculpting is verified in \cref{sec:sculpting}.
The overall anomaly detection performance and the sensitivity of the new method are evaluated in \cref{sec:sensitivity}.

The \gls{NF} architecture and definitions of \gls{SR} and \gls{SB} follow the configuration of Ref.~\cite{haxadv1}.
For the weakly supervised classifier, a similar strategy is adopted with minor modifications.
Each classifier is a \gls{BDT} trained with 5-fold cross-validation to prevent overfitting and make efficient use of the available data.
To reduce fluctuations from limited training statistics and model initialization, five independent encoders are trained, each paired with four NFs that together generate the background estimate.
For each of the five encoder--NF combinations, four BDTs are trained, each with 5-fold cross-validation, yielding 20 classifiers in total.
These are randomly grouped into ten ensembles of two, and the scores within each ensemble are averaged.
Compared to Ref.~\cite{haxadv1}, the ensemble size is reduced from four to two BDTs, which helps suppress background sculpting.
The spread across the ten ensembles is used to estimate the systematic uncertainty of the initialization and training of the ML models.
The default classifier working point is tightened from $\varepsilon_B = 0.5\%$ to $\varepsilon_B = 0.1\%$, retaining only the most signal-like fraction of events in the \gls{SR} and thereby improving the sensitivity.
The $\varepsilon_B = 0.5\%$ working point of Ref.~\cite{haxadv1} is retained for direct comparison.
This choice is further motivated by the results in \cref{sec:in_dataset}.

Since this study is performed on simulated data, it is important to carefully assess how discrepancies between simulated data and recorded data could affect our results.
The encoder training uses simulated data and is sensitive to data-MC differences, as these can distort how events are embedded into the learned latent space.
This effect could be mitigated by incorporating control-region data into the encoder training or by imposing stronger regularization, which we leave to future work.
The CATHODE method, by contrast, remains fully data-driven even when operating on the simulation-informed embedding space, and is therefore largely unaffected by such discrepancies.

\section{Results}
\label{sec:results}

This section reviews the results of the method in three steps:
First, the structure of the contrastive latent space is examined to verify that it satisfies the design requirements of signal sensitivity and modelability (\cref{sec:latent_space}).
Second, the classifier performance is studied to confirm the absence of background sculpting in the $m_{\gamma\gamma}$ spectrum.
Third, the anomaly detection sensitivity is exhaustively examined and quantified for the different training configurations, including a comparison to existing methods.

\subsection{Latent Space}
\label{sec:latent_space}
\begin{figure*}[htbp]
  \centering
  \subfloat[]{\includegraphics[width=0.32\textwidth,page=1]{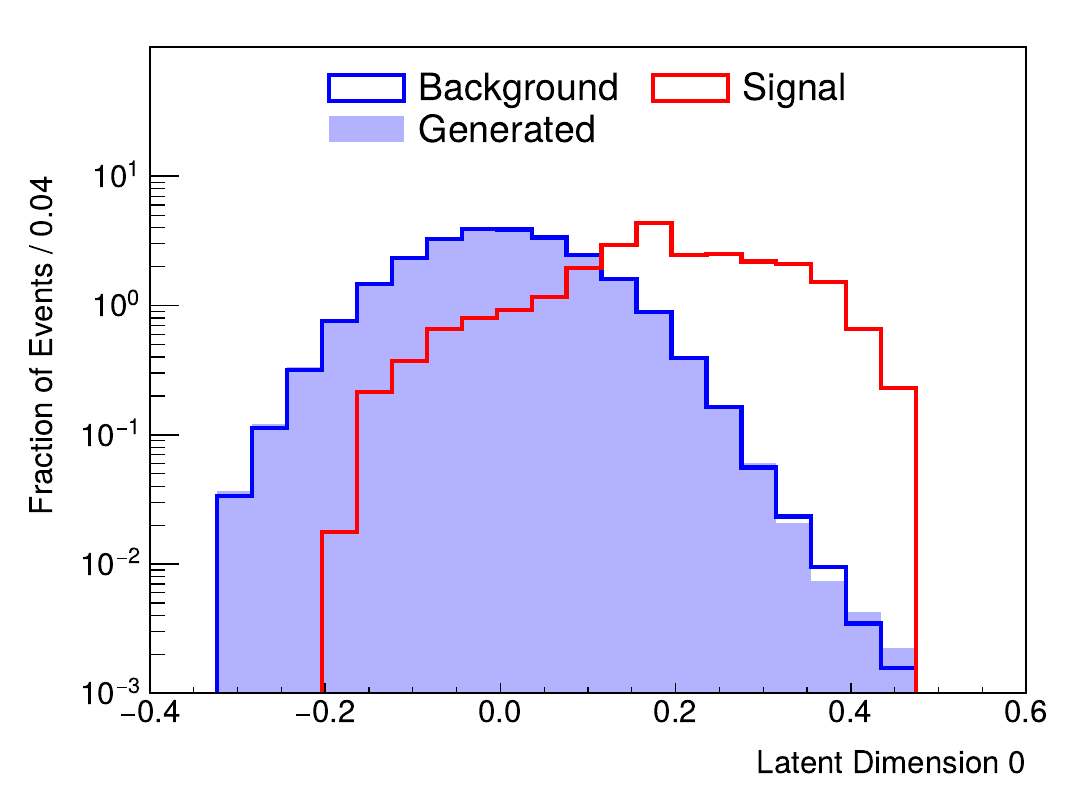}\label{fig:feat_1}}
  \hfill
  \subfloat[]{\includegraphics[width=0.32\textwidth,page=2]{latent_space_plots/feature_distribution}\label{fig:feat_2}}
  \hfill
  \subfloat[]{\includegraphics[width=0.32\textwidth,page=3]{latent_space_plots/feature_distribution}\label{fig:feat_3}}\\
  \subfloat[]{\includegraphics[width=0.32\textwidth,page=4]{latent_space_plots/feature_distribution}\label{fig:feat_4}}
  \hfill
  \subfloat[]{\includegraphics[width=0.32\textwidth,page=5]{latent_space_plots/feature_distribution}\label{fig:feat_5}}
  \hfill
  \subfloat[]{\includegraphics[width=0.32\textwidth,page=6]{latent_space_plots/feature_distribution}\label{fig:feat_6}}
  \caption{One-dimensional projections of the six latent space features for the \gls{ID} configuration using the pseudo-data in the \gls{SR}. The distribution of the background from the pseudo-data (blue outline), the generated background from the \gls{NF} (blue filled), and $(\chi_1^{\pm}\chi_2^{0})_{200}$ signal (red) are compared. Background refers to both the non-resonant and resonant \gls{SM} Higgs components.
  }
  \label{fig:feature_distribution}
\end{figure*}

\Cref{fig:feature_distribution} shows the one-dimensional projections of the six latent space features for the \gls{ID} configuration using the pseudo data in the \gls{SR}.
The background distributions are smooth and approximately Gaussian, confirming that the \gls{KL} regularization produces a latent space that is well suited for density estimation.
Even in these one-dimensional projections, the tested signal is visibly separated from the background in most latent dimensions, demonstrating that the contrastive training encodes discriminating information across the full latent space.
The generated background, produced by the \gls{NF} trained on sideband data, accurately reproduces the true background distributions in all six dimensions, validating the downstream density estimation.

\begin{figure*}[htbp]
  \centering
  \subfloat[All processes (in dataset)]{\includegraphics[width=0.32\textwidth]{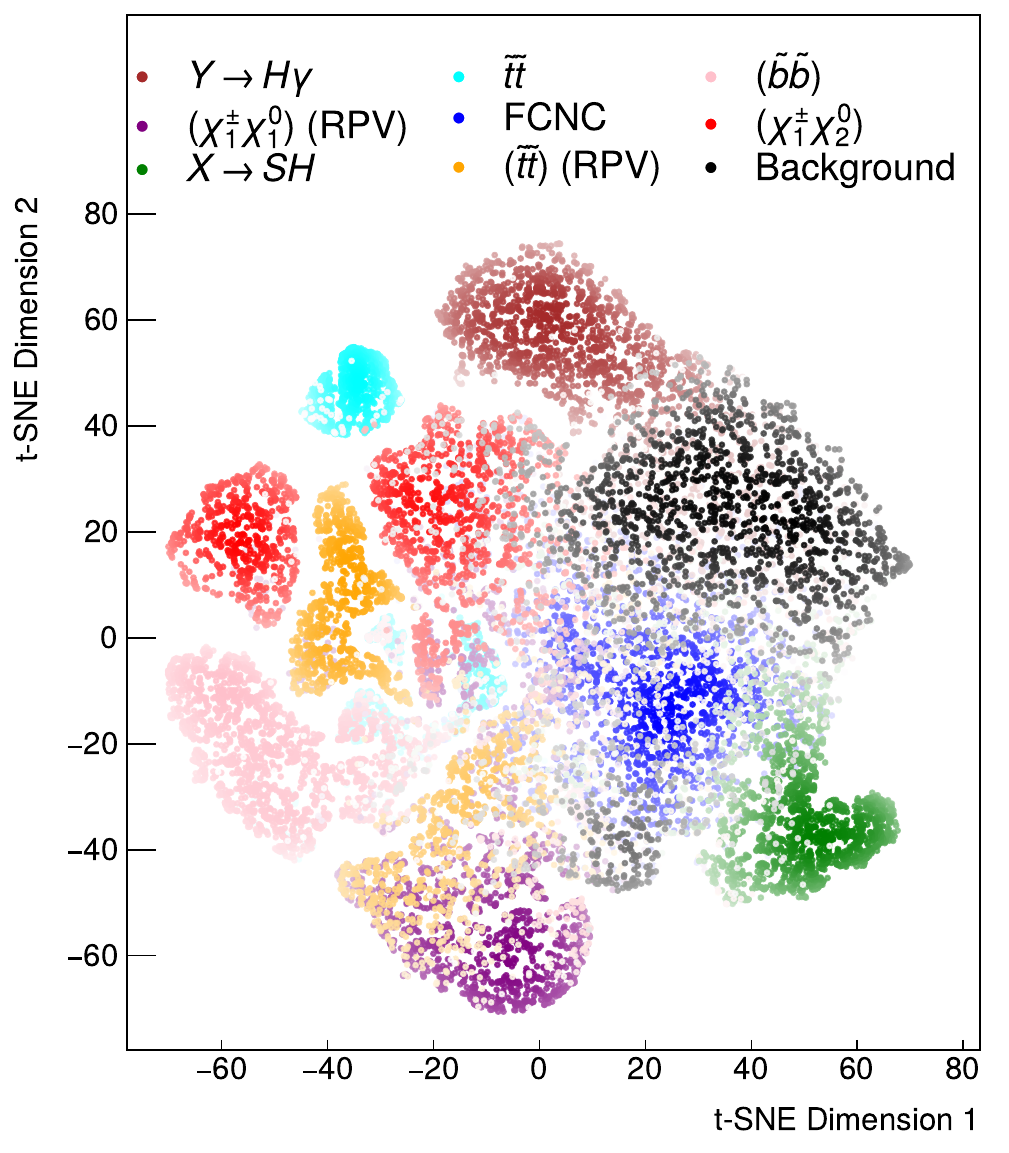}\label{fig:tsne_all}}
  \hfill
  \subfloat[Interpolation holdout]{\includegraphics[width=0.32\textwidth]{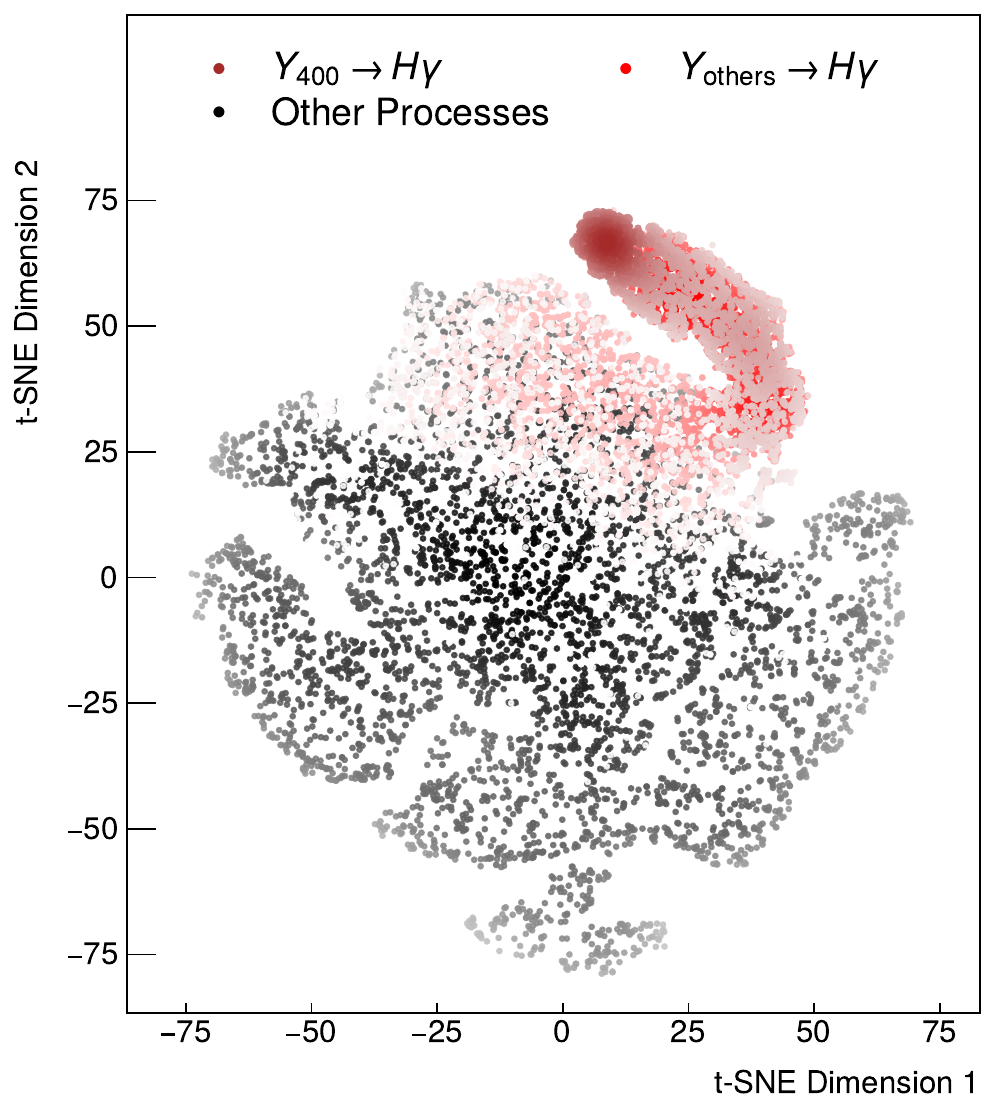}\label{fig:tsne_holdout_zp}}
  \hfill
  \subfloat[Background only]{\includegraphics[width=0.32\textwidth]{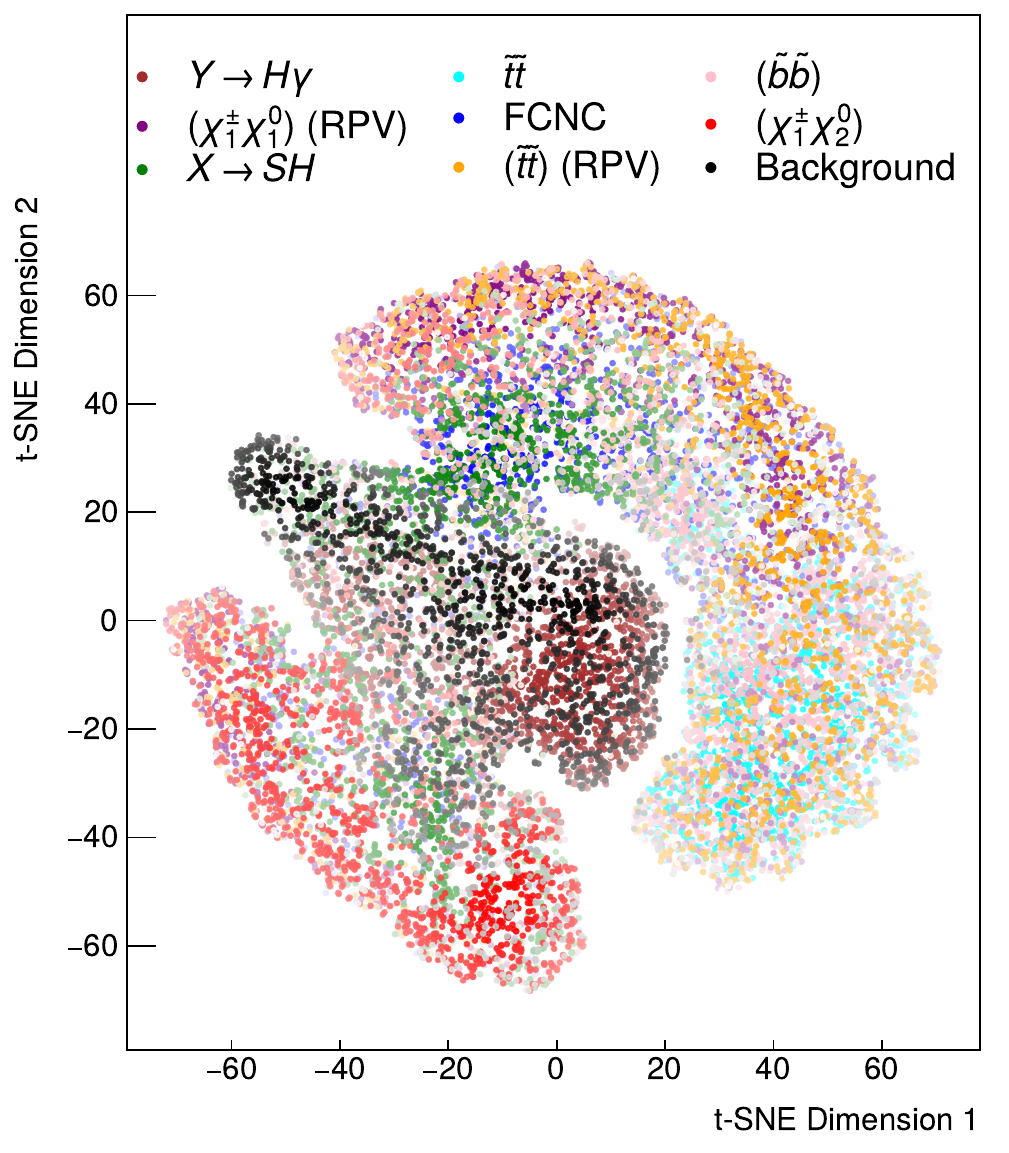}\label{fig:tsne_bkg}}
  \caption{$t$-SNE visualizations of the contrastive latent space for three embedding configurations: \protect\subref{fig:tsne_all} trained with all signal and background processes (\gls{ID}), \protect\subref{fig:tsne_holdout_zp} with the $Y_{400} \to H\gamma$ mass point held out for interpolation (\gls{IP}), and \protect\subref{fig:tsne_bkg} trained with background processes only. In \protect\subref{fig:tsne_all} and \protect\subref{fig:tsne_bkg} ``Background'' refers to SM non-resonant background and SM Higgs processes, in \protect\subref{fig:tsne_holdout_zp} ``Other Processes'' includes all physics processes other than $Y_m \to H\gamma$.}
  \label{fig:tsne_embeddings}
\end{figure*}

\Cref{fig:tsne_embeddings} shows $t$-SNE visualizations~\cite{JMLR:v9:vandermaaten08a} of the latent space for three training configurations.
In the \gls{ID} case (\cref{fig:tsne_all}), the \gls{SM} backgrounds and \gls{BSM} signals form well-separated clusters, reflecting the contrastive objective.
In the \gls{IP} case (holding out $Y_{400} \to H\gamma$ in \cref{fig:tsne_holdout_zp}), the held-out signal still occupies a distinct region of the latent space rather than collapsing onto the background.
Additionally, the events of the held-out signal are located adjacent to the cluster of the same physics process, showing how the encoder generalizes to new parametrizations of known signal topologies.
In the background-only configuration (\cref{fig:tsne_bkg}), the separation between processes is significantly reduced, in particular between \gls{SM} backgrounds and \gls{BSM} signals, underscoring the benefit of the inclusion of signal models in the contrastive training.
The information preserved in the latent space is evaluated in \cref{app:iad_comparisons}.

\subsection{Sculpting Check}
\label{sec:sculpting}

To verify that the method does not introduce a signal-like signature in the $m_{\gamma\gamma}$ spectrum in the absence of a signal in the data, a dedicated sculpting test is performed.
The test uses the \gls{ID} configuration, and whereas the training of the contrastive latent space uses all signal processes, no signal is injected into the data in the \gls{CATHODE} stage.
A classifier is trained to distinguish the \gls{NF}-generated background from the true background events in the \gls{SR}.
If the generated and true backgrounds differ, the classifier will learn to exploit this difference; conversely, an \gls{AUC} of 0.5 indicates that the two samples are indistinguishable.

\begin{figure}[htbp]
  \centering
  \subfloat[Classifier ROC curve]{\includegraphics[width=\columnwidth,trim={0.45cm 0 0 0}]{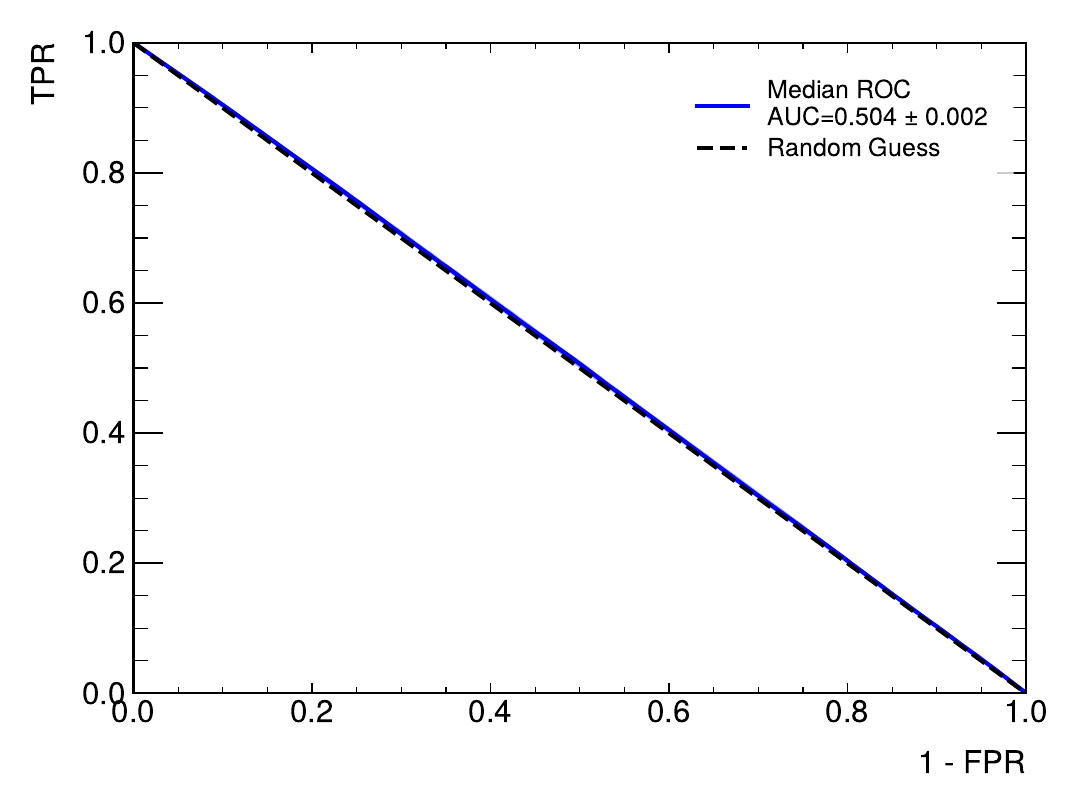}\label{fig:sculpting_roc}}\\
  \subfloat[$m_{\gamma\gamma}$ spectrum]{\includegraphics[width=\columnwidth,trim={0 0 0.1cm 0}]{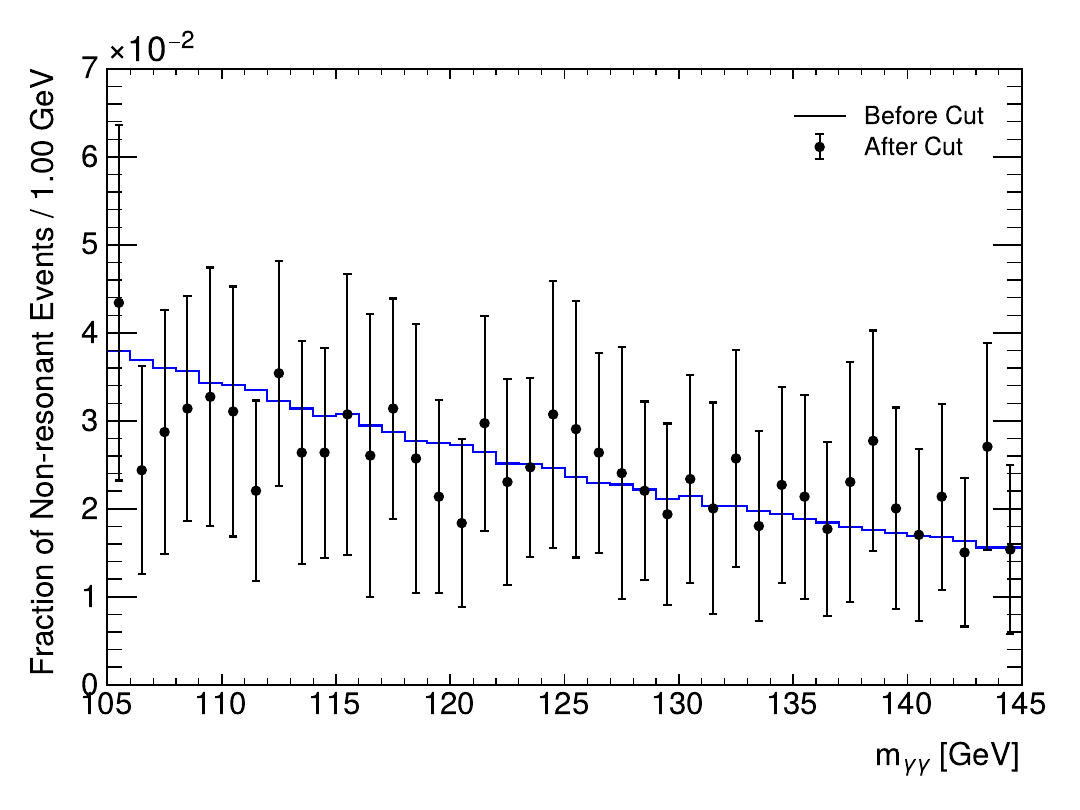}\label{fig:myy_shape}}
  \caption{Sculpting tests for the \gls{ID} configuration without signal injection. \protect\subref{fig:sculpting_roc} \gls{ROC} curve of a classifier trained to distinguish \gls{NF}-generated background from true background in the \gls{SR}, yielding an \gls{AUC} of 0.504 $\pm$ 0.002. \protect\subref{fig:myy_shape} Normalized $m_{\gamma\gamma}$ spectrum before (blue) and after (black) a cut on the classifier output. The uncertainties include statistical effects and uncertainties from the initialization and training of the ML models.}
  \label{fig:sculpting}
\end{figure}

\Cref{fig:sculpting_roc} shows the \gls{ROC} curve of this classification, yielding an \gls{AUC} of 0.504 $\pm$ 0.002, indicating near-random discrimination between the two samples.
This confirms that the \gls{NF} density estimation in the regularized latent space produces a background estimate that is faithful to the true background.
\Cref{fig:myy_shape} shows the normalized $m_{\gamma\gamma}$ spectrum before and after applying the cut on the classifier output.
The two distributions agree within their statistical uncertainties, confirming that the method does not significantly change the shape of the $m_{\gamma\gamma}$ spectrum and, most importantly, does not produce a spurious bump that could mimic a signal.
The \gls{ID} configuration is the most challenging case to regularize, as it includes the largest number of process groups.
The fact that no sculpting is observed in this configuration provides confidence that the other configurations are equally well behaved.

\subsection{Sensitivity}
\label{sec:sensitivity}

The sensitivity of the method is evaluated separately for each of the different embedding configurations.
The \gls{SIC} is used to evaluate the performance of the classifiers as a function of the injected signal strength.
In all following tests, for each signal model at each injection strength, 10 different random instances of the initial datasets are used to estimate statistical effects.

\subsubsection{In Dataset}
\label{sec:in_dataset}

\begin{figure*}[htbp]
  \centering
  \subfloat[$(\chi_1^{\pm}\chi_2^{0})_{150}$]{\includegraphics[width=0.48\textwidth,page=1]{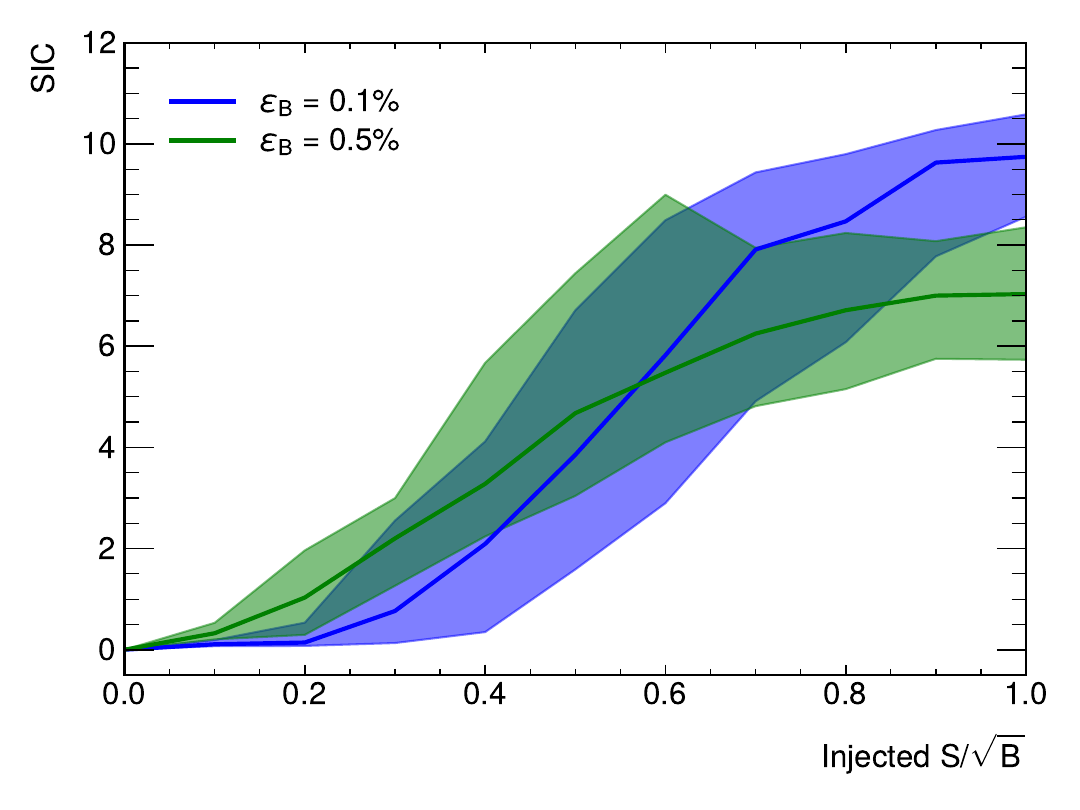}\label{fig:sic_haxad_wnh150}}
  \hfill
  \subfloat[$X \to S(\ell\ell)H$]{\includegraphics[width=0.48\textwidth,page=2]{performance_plots/sic_scan_haxad_comparison}\label{fig:sic_haxad_xsh}}
  \caption{\gls{SIC} as a function of the injected signal strength for the \gls{ID} configuration. Two classifier working points are shown: $\varepsilon_B = 0.5\%$ (matching the working point of Ref.~\cite{haxadv1}) and $\varepsilon_B = 0.1\%$. \protect\subref{fig:sic_haxad_wnh150} Chargino--neutralino signal $(\chi_1^{\pm}\chi_2^{0})_{150}$. \protect\subref{fig:sic_haxad_xsh} Extended Higgs sector signal $X \to S(\ell\ell)H$. The uncertainty bands include statistical effects and uncertainties from the initialization and training of the ML models.}
  \label{fig:sic_haxad_comparison}
\end{figure*}

The \gls{ID} configuration uses all signal models during the contrastive training and serves as the point of reference for comparison with Ref.~\cite{haxadv1}, which uses the same dataset and analysis strategy but with a different embedding approach.
\Cref{fig:sic_haxad_comparison} shows the \gls{SIC} as a function of the injected signal strength for two representative signal models at two classifier working points ($\varepsilon_B = 0.5\%$ and $\varepsilon_B = 0.1\%$).
At a signal strength corresponding to a $1\,\sigma$ injection of the chargino--neutralino signal $(\chi_1^{\pm}\chi_2^{0})_{150}$ (\cref{fig:sic_haxad_wnh150}), the contrastive embedding achieves an \gls{SIC} of approximately 7.0 at the $\varepsilon_B = 0.5\%$ working point used in Ref.~\cite{haxadv1}, compared to a value of 5.3 obtained in that work.
At the tighter $\varepsilon_B = 0.1\%$ working point, the \gls{SIC} reaches 9.8.
For the extended Higgs sector signal $X \to S(\ell\ell)H$ (\cref{fig:sic_haxad_xsh}), the improvement is even more pronounced: the \gls{SIC} increases from 6.4 in Ref.~\cite{haxadv1} to 9.0, and reaches 17.8 at the tighter working point.
The stronger cut is enabled by the improved regularization of the contrastive latent space, which allows more aggressive signal selection without introducing background sculpting.
Overall, the contrastive embedding in the \gls{ID} configuration improves the \gls{SIC} by around 40\% on the two signal models presented in Ref.~\cite{haxadv1}.

While the \gls{ID} configuration offers the highest sensitivity, it requires the tested signal model to be included in the embedding training, and therefore does not constitute an entirely model-agnostic anomaly detection strategy.
The sensitivity of the \gls{ID} configuration across all tested signal models and configurations is summarized in \cref{sec:summary}.

\subsubsection{Interpolation}
\label{sec:interpolation}

The \gls{IP} configuration uses a contrastive embedding trained on all signal topologies, but with specific parametrizations held out.
For instance, the heavy neutral resonance process $Y_m \to H\gamma$ is included in training at masses of 200, 300, and \SI{500}{\GeV}, but the \SI{400}{\GeV} point is withheld and used only for evaluation.
This tests whether the embedding generalizes across the parameter space of a known signal topology.

\begin{figure*}[htbp]
  \centering
  \subfloat[$Y_{400} \to H\gamma$]{\includegraphics[width=0.48\textwidth,page=1]{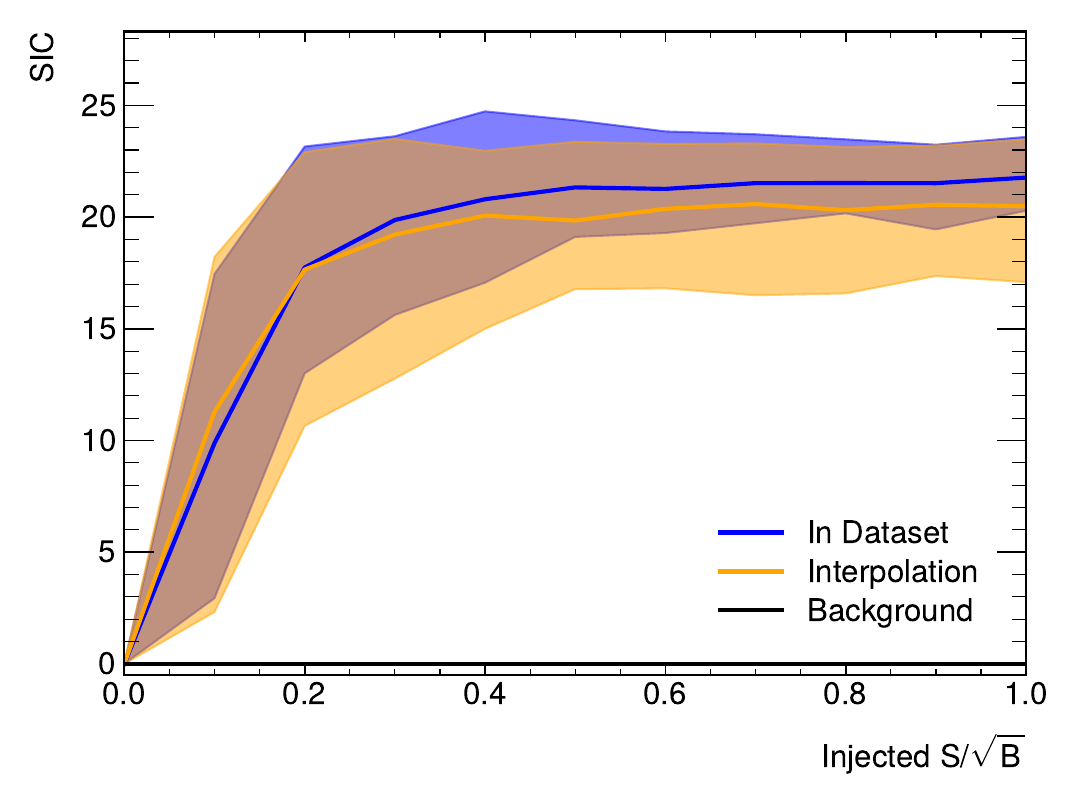}\label{fig:sic_ip_zphyy}}
  \hfill
  \subfloat[$(\chi_1^{\pm}\chi_2^{0})_{200}$]{\includegraphics[width=0.48\textwidth,page=2]{performance_plots/sic_scan_param_interpolation}\label{fig:sic_ip_wnh200}}
  \caption{\gls{SIC} as a function of the injected signal strength for \gls{IP} signals not included in the embedding training. Results are shown for the \gls{ID}, \gls{IP}, and background-only configurations. \protect\subref{fig:sic_ip_zphyy} Heavy neutral resonance $Y_{400} \to H\gamma$. \protect\subref{fig:sic_ip_wnh200} Chargino--neutralino $(\chi_1^{\pm}\chi_2^{0})_{200}$. The uncertainty bands include statistical effects and uncertainties from the initialization and training of the ML models.}
  \label{fig:sic_interpolation}
\end{figure*}

\Cref{fig:sic_interpolation} compares the \gls{SIC} for the \gls{ID}, \gls{IP}, and background-only configurations for two held-out parametrizations.
For the heavy neutral resonance $Y_{400} \to H\gamma$ (\cref{fig:sic_ip_zphyy}), the \gls{IP} configuration nearly matches the \gls{ID} performance, while the background-only embedding fails to pick up the signal entirely.
For the chargino--neutralino $(\chi_1^{\pm}\chi_2^{0})_{200}$ (\cref{fig:sic_ip_wnh200}), the \gls{IP} configuration again approaches the \gls{ID} performance while substantially outperforming the background-only embedding.

The \gls{IP} configuration is more general than the \gls{ID} case, as it does not require the exact signal parametrization to be present in the training data.
However, it still requires the signal topology to be included, limiting its applicability to parametric extensions of hypothesized processes.

\subsubsection{Extrapolation}
\label{sec:extrapolation}

\begin{figure*}[htbp]
  \centering
  \subfloat[$(\chi_1^{\pm}\chi_1^{0})_{200}$ (RPV)]{\includegraphics[width=0.48\textwidth,page=1]{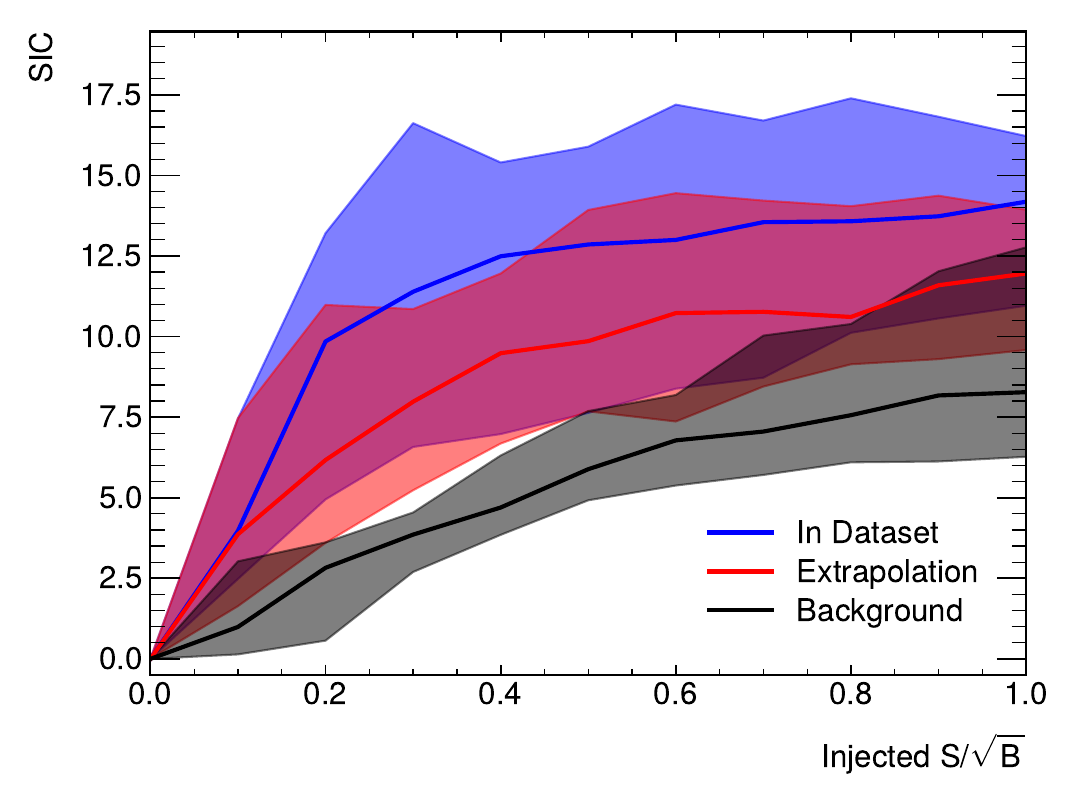}\label{fig:sic_ep_rpv}}
  \hfill
  \subfloat[$\tilde{t}\,\tilde{t}$]{\includegraphics[width=0.48\textwidth,page=2]{performance_plots/sic_scan_model_extrapolation}\label{fig:sic_ep_stop}}
  \caption{\gls{SIC} as a function of the injected signal strength for \gls{EP} signals from model classes not included in the embedding training. Results are shown for the \gls{ID}, \gls{EP}, and background-only configurations. \protect\subref{fig:sic_ep_rpv} RPV neutralino $(\chi_1^{\pm}\chi_1^{0})_{200}$. \protect\subref{fig:sic_ep_stop} RPC stop $\tilde{t}\,\tilde{t}$. The uncertainty bands include statistical effects and uncertainties from the initialization and training of the ML models.}
  \label{fig:sic_extrapolation}
\end{figure*}

The \gls{EP} configuration holds out an entire signal topology from the contrastive training, testing whether the embedding generalizes to previously unseen classes of \gls{BSM} physics.
\Cref{fig:sic_extrapolation} shows the \gls{SIC} for the RPV neutralino $(\chi_1^{\pm}\chi_1^{0})_{200}$ and the RPC stop $\tilde{t}\,\tilde{t}$, comparing the \gls{ID}, \gls{EP}, and background-only configurations.
In both cases, the \gls{EP} configuration outperforms the background-only embedding, demonstrating that including a diverse set of \gls{BSM} signals in the contrastive training improves the latent space even for signal topologies not present in the training data.

The \gls{EP} configuration represents the most general signal-aware setup, as it requires no prior knowledge of the specific signal topology under test.
A quantitative comparison of the \gls{EP}, \gls{ID}, and background-only configurations across all signal models is presented in the following section.

\subsubsection{Summary}
\label{sec:summary}

\begin{figure*}[htbp]
  \centering
  \includegraphics[width=\textwidth]{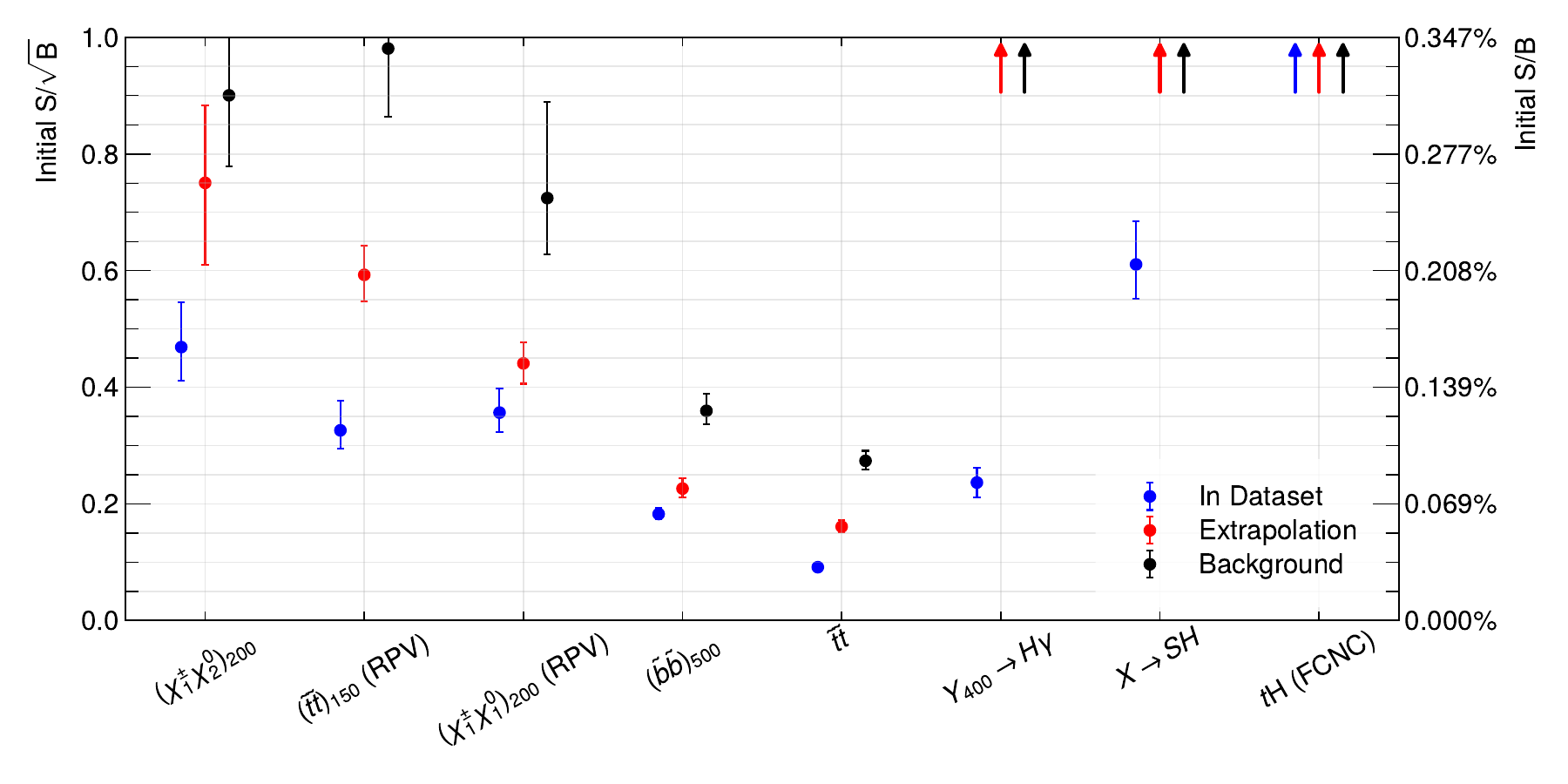}
  \caption{Summary of the anomaly detection sensitivity across all signal models. Shown is the initial pre-selection $S/\sqrt{B}$ required to achieve a post-selection significance of $3\sigma$, for the \gls{ID}, \gls{EP}, and background-only embedding configurations. Upward arrows indicate that $3\sigma$ is not reached within the scanned injection range. The uncertainty bars include statistical effects and uncertainties from the initialization and training of the ML models.}
  \label{fig:comparison}
\end{figure*}

To compare the sensitivity across all signal models and embedding configurations, \cref{fig:comparison} shows the initial pre-selection signal injection, expressed as $S/\sqrt{B}$, required to achieve a post-selection significance of $3\sigma$ after the full \gls{CATHODE} pipeline.
Lower values indicate higher sensitivity, and upward arrows denote configurations that do not reach $3\sigma$ within the scanned injection range.

The \gls{ID} configuration consistently achieves the lowest required injection across all signal models, confirming that full knowledge of the signal topology during embedding training yields the highest sensitivity.
The \gls{EP} configuration generally requires a higher injection than \gls{ID} but outperforms the background-only embedding for the majority of tested signals, reinforcing the finding that including diverse \gls{BSM} signals in the contrastive training improves sensitivity even to unseen topologies.

For the three signal models on the right of \cref{fig:comparison}, neither the \gls{EP} nor the background-only configuration reaches $3\sigma$ within the scanned range.
Among these, the \gls{ID} configuration is the only one that reaches $3\sigma$ for some of these signals, while for $tH$ (FCNC) even \gls{ID} falls short.
These cases highlight the limits of the contrastive embedding approach and motivate further developments.

\section{Conclusion}
\label{sec:conclusion}

This work presents a signal-aware supervised contrastive learning approach for constructing a low-dimensional latent space to advance weakly supervised anomaly detection.
A particle transformer encoder is trained with a combined loss function that uses supervised contrastive learning to separate physics processes in the latent space while simultaneously regularizing the latent distributions toward a unit Gaussian prior through a KL divergence term.
The resulting latent space is both signal-sensitive and well suited for modeling by generative models.
The method was tested in a diphoton final state using eight \gls{BSM} signal groups spanning SUSY scenarios, extended Higgs sectors, heavy neutral resonances, and flavor-changing neutral currents.
Four embedding configurations were compared: \gls{ID}, which includes all signal models in training; \gls{IP}, which holds out specific signal parametrizations; \gls{EP}, which holds out entire signal topologies; and a background-only baseline trained without any \gls{BSM} signals.

The signal-aware contrastive latent space proves highly effective for weakly supervised anomaly detection.
In the \gls{ID} configuration, the method improves the \gls{SIC} by around 40\% on the two signal models presented in Ref.~\cite{haxadv1}, which uses a simpler approach based on a \gls{VAE}. 
It also uses a much broader set of input features than Ref.~\cite{haxadv1}, which potentially increases sensitivity to new physics over a larger region of phase space.
The improved regularization enables tighter classifier working points, further increasing the sensitivity.
The \gls{IP} configuration nearly matches the \gls{ID} performance for held-out signal parametrizations, demonstrating that the embedding generalizes across the parameter space of known signal topologies.
The \gls{EP} configuration outperforms the background-only embedding for unseen signal topologies, showing that the inclusion of diverse \gls{BSM} signals in the contrastive training improves sensitivity even to new physics classes not present during training.
No background sculpting is observed in the $m_{\gamma\gamma}$ spectrum, even in the most challenging configuration.

By bridging supervised latent space construction with weakly supervised anomaly detection, this approach offers a viable path toward high-dimensional anomaly detection at the LHC and beyond.
A natural next step is the application of this method to recorded collision data.

\section*{Code and Data Availability}

The code for this project is available at: \url{https://github.com/Nollde/contrast_embed}.
The datasets used in this study are available upon request to the authors.

\section*{Acknowledgments}
We thank Qibin Liu for his help with the preparation of the signal samples.
We thank Sascha Diefenbacher, Chi Lung Cheng, and Qibin Liu for valuable comments on the manuscript.
B.N. and D.N. are supported by the Department of Energy (DOE), Office of Science under contract DE-AC02-76SF00515 and R.L. is supported by DOE grant DE-SC0017660.
This research used resources of the National Energy Research Scientific Computing Center, a DOE Office of Science User Facility supported by the Office of Science of the U.S. Department of Energy under Contract No. DE-AC02-05CH11231 using NERSC award HEP-ERCAP0035546.

\bibliography{HEPML,HEPML_add}

@misc{adam,
  title={{Adam}: A method for stochastic optimization},
  author={Kingma, Diederik and Ba, Jimmy},
  eprint = "1412.6980",
  archivePrefix  = "arXiv",
  primaryClass = "cs",
  year={2014}
}

@misc{kingma2014autoencoding,
      title={{Auto}-{Encoding} {Variational} {Bayes}},
      author={Diederik P Kingma and Max Welling},
      year={2014},
      eprint={1312.6114},
      archivePrefix={arXiv},
      primaryClass={stat.ML}
}

@article{Cacciari:2011ma,
    author = "Cacciari, Matteo and Salam, Gavin P. and Soyez, Gregory",
    title = "{FastJet User Manual}",
    eprint = "1111.6097",
    archivePrefix = "arXiv",
    primaryClass = "hep-ph",
    reportNumber = "CERN-PH-TH-2011-297",
    doi = "10.1140/epjc/s10052-012-1896-2",
    journal = "Eur. Phys. J. C",
    volume = "72",
    pages = "1896",
    year = "2012"
}

@article{Cacciari:2008gp,
    author = "Cacciari, Matteo and Salam, Gavin P. and Soyez, Gregory",
    title = "{The anti-$k_t$ jet clustering algorithm}",
    eprint = "0802.1189",
    archivePrefix = "arXiv",
    primaryClass = "hep-ph",
    reportNumber = "LPTHE-07-03",
    doi = "10.1088/1126-6708/2008/04/063",
    journal = "JHEP",
    volume = "04",
    pages = "063",
    year = "2008"
}

@article{Cacciari:2005hq,
    author = "Cacciari, Matteo and Salam, Gavin P.",
    title = "{Dispelling the $N^{3}$ myth for the $k_t$ jet-finder}",
    eprint = "hep-ph/0512210",
    archivePrefix = "arXiv",
    reportNumber = "LPTHE-05-32",
    doi = "10.1016/j.physletb.2006.08.037",
    journal = "Phys. Lett. B",
    volume = "641",
    pages = "57--61",
    year = "2006"
}

@article{Karagiorgi:2021ngt,
    author = "Karagiorgi, Georgia and Kasieczka, Gregor and Kravitz, Scott and Nachman, Benjamin and Shih, David",
    title = "{Machine Learning in the Search for New Fundamental Physics}",
    eprint = "2112.03769",
    archivePrefix = "arXiv",
    primaryClass = "hep-ph",
    month = "12",
    year = "2021",
    doi = "10.1038/s42254-022-00455-1",
    journal = "Nature Reviews Physics",
    volume = "4",
    number = "6",
    pages = "399--412",
}

@article{Hallin:2021wme,
    author = "Hallin, Anna and Isaacson, Joshua and Kasieczka, Gregor and Krause, Claudius and Nachman, Benjamin and Quadfasel, Tobias and Schlaffer, Matthias and Shih, David and Sommerhalder, Manuel",
    title = "{Classifying anomalies through outer density estimation}",
    eprint = "2109.00546",
    archivePrefix = "arXiv",
    primaryClass = "hep-ph",
    reportNumber = "EFI-20-5, FERMILAB-PUB-21-389-T",
    doi = "10.1103/PhysRevD.106.055006",
    journal = "Phys. Rev. D",
    volume = "106",
    number = "5",
    pages = "055006",
    year = "2022"
}

@article{Kasieczka:2021xcg,
    author = "Kasieczka, Gregor and others",
    title = "{The LHC Olympics 2020 a community challenge for anomaly detection in high energy physics}",
    eprint = "2101.08320",
    archivePrefix = "arXiv",
    primaryClass = "hep-ph",
    doi = "10.1088/1361-6633/ac36b9",
    journal = "Rept. Prog. Phys.",
    volume = "84",
    number = "12",
    pages = "124201",
    year = "2021"
}

@article{ATLAS:2020iwa,
    author = "{ATLAS Collaboration}",
    title = "{Dijet resonance search with weak supervision using $\sqrt{s}=13$ TeV $pp$ collisions in the ATLAS detector}",
    eprint = "2005.02983",
    archivePrefix = "arXiv",
    primaryClass = "hep-ex",
    reportNumber = "CERN-EP-2020-062",
    doi = "10.1103/PhysRevLett.125.131801",
    journal = "Phys. Rev. Lett.",
    volume = "125",
    number = "13",
    pages = "131801",
    year = "2020"
}

@article{Nachman:2020lpy,
    author = "Nachman, Benjamin and Shih, David",
    title = "{Anomaly Detection with Density Estimation}",
    eprint = "2001.04990",
    archivePrefix = "arXiv",
    primaryClass = "hep-ph",
    doi = "10.1103/PhysRevD.101.075042",
    journal = "Phys. Rev. D",
    volume = "101",
    pages = "075042",
    year = "2020"
}

@article{Collins:2018epr,
      author         = "Collins, Jack H. and Howe, Kiel and Nachman, Benjamin",
      title          = "{Anomaly Detection for Resonant New Physics with Machine Learning}",
      journal        = "Phys. Rev. Lett.",
      volume         = "121",
      year           = "2018",
      number         = "24",
      pages          = "241803",
      doi            = "10.1103/PhysRevLett.121.241803",
      eprint         = "1805.02664",
      archivePrefix  = "arXiv",
      primaryClass   = "hep-ph",
      reportNumber   = "FERMILAB-PUB-18-180-T",
      SLACcitation   = "%%CITATION = ARXIV:1805.02664;%%"
}

@article{Metodiev:2017vrx,
    author = "Metodiev, Eric M. and Nachman, Benjamin and Thaler, Jesse",
    title = "{Classification without labels: Learning from mixed samples in high energy physics}",
    eprint = "1708.02949",
    archivePrefix = "arXiv",
    primaryClass = "hep-ph",
    reportNumber = "MIT--CTP-4922",
    doi = "10.1007/JHEP10(2017)174",
    journal = "JHEP",
    volume = "10",
    pages = "174",
    year = "2017"
}

@article{ATLAS:2025obc,
    author = "{ATLAS Collaboration}",
    title = "{Weakly supervised anomaly detection for resonant new physics in the dijet final state using proton-proton collisions at $\sqrt{s}=13$ TeV with the ATLAS detector}",
    eprint = "2502.09770",
    archivePrefix = "arXiv",
    primaryClass = "hep-ex",
    reportNumber = "CERN-EP-2025-002",
    doi = "10.1103/2yq5-vj59",
    journal = "Phys. Rev. D",
    volume = "112",
    number = "7",
    pages = "072009",
    year = "2025"
}

@article{Gambhir:2025afb,
    author = "Gambhir, Rikab and Mastandrea, Radha and Nachman, Benjamin and Thaler, Jesse",
    title = "{Isolating Unisolated Upsilons with Anomaly Detection in CMS Open Data}",
    eprint = "2502.14036",
    archivePrefix = "arXiv",
    primaryClass = "hep-ph",
    reportNumber = "MIT-CTP 5843",
    doi = "10.1103/vvv3-5kkl",
    journal = "Phys. Rev. Lett.",
    volume = "135",
    number = "2",
    pages = "021902",
    year = "2025"
}

@article{CMS:2024nsz,
    author = "{CMS Collaboration}",
    title = "{Model-agnostic search for dijet resonances with anomalous jet substructure in proton{\textendash}proton collisions at $\sqrt{s}$ = 13 TeV}",
    eprint = "2412.03747",
    archivePrefix = "arXiv",
    primaryClass = "hep-ex",
    reportNumber = "CMS-EXO-22-026, CERN-EP-2024-291",
    doi = "10.1088/1361-6633/add762",
    journal = "Rept. Prog. Phys.",
    volume = "88",
    number = "6",
    pages = "067802",
    year = "2025"
}

@article{Belis:2023mqs,
    author = "Belis, Vasilis and Odagiu, Patrick and Aarrestad, Thea Klaeboe",
    title = "{Machine learning for anomaly detection in particle physics}",
    eprint = "2312.14190",
    archivePrefix = "arXiv",
    primaryClass = "physics.data-an",
    doi = "10.1016/j.revip.2024.100091",
    journal = "Rev. Phys.",
    volume = "12",
    pages = "100091",
    year = "2024"
}

@article{Sjostrand:2006za,
    author = "Sjostrand, Torbjorn and Mrenna, Stephen and Skands, Peter Z.",
    title = "{PYTHIA 6.4 Physics and Manual}",
    eprint = "hep-ph/0603175",
    archivePrefix = "arXiv",
    reportNumber = "FERMILAB-PUB-06-052-CD-T, LU-TP-06-13",
    doi = "10.1088/1126-6708/2006/05/026",
    journal = "JHEP",
    volume = "05",
    pages = "026",
    year = "2006"
}

@article{Sjostrand:2014zea,
    author = {Sj{\"o}strand, Torbj{\"o}rn and Ask, Stefan and Christiansen, Jesper R. and Corke, Richard and Desai, Nishita and Ilten, Philip and Mrenna, Stephen and Prestel, Stefan and Rasmussen, Christine O. and Skands, Peter Z.},
    title = "{An introduction to PYTHIA 8.2}",
    eprint = "1410.3012",
    archivePrefix = "arXiv",
    primaryClass = "hep-ph",
    reportNumber = "LU-TP-14-36, MCNET-14-22, CERN-PH-TH-2014-190, FERMILAB-PUB-14-316-CD, DESY-14-178, SLAC-PUB-16122",
    doi = "10.1016/j.cpc.2015.01.024",
    journal = "Comput. Phys. Commun.",
    volume = "191",
    pages = "159--177",
    year = "2015"
}

@article{deFavereau:2013fsa,
    author = "{The DELPHES 3 collaboration}",
    title = "{DELPHES 3, A modular framework for fast simulation of a generic collider experiment}",
    eprint = "1307.6346",
    archivePrefix = "arXiv",
    primaryClass = "hep-ex",
    doi = "10.1007/JHEP02(2014)057",
    journal = "JHEP",
    volume = "02",
    pages = "057",
    year = "2014"
}

@article{Mertens:2015kba,
    author = "Mertens, Alexandre",
    editor = "Fiala, L. and Lokajicek, M. and Tumova, N.",
    title = "{New features in Delphes 3}",
    doi = "10.1088/1742-6596/608/1/012045",
    journal = "J. Phys. Conf. Ser.",
    volume = "608",
    number = "1",
    pages = "012045",
    year = "2015"
}

@article{Hallin:2022eoq,
    author = "Hallin, Anna and Kasieczka, Gregor and Quadfasel, Tobias and Shih, David and Sommerhalder, Manuel",
    title = "{Resonant anomaly detection without background sculpting}",
    eprint = "2210.14924",
    archivePrefix = "arXiv",
    primaryClass = "hep-ph",
    doi = "10.1103/PhysRevD.107.114012",
    journal = "Phys. Rev. D",
    volume = "107",
    number = "11",
    pages = "114012",
    year = "2023"
}

@article{Freytsis:2023cjr,
    author = "Freytsis, Marat and Perelstein, Maxim and San, Yik Chuen",
    title = "{Anomaly detection in the presence of irrelevant features}",
    eprint = "2310.13057",
    archivePrefix = "arXiv",
    primaryClass = "hep-ph",
    doi = "10.1007/JHEP02(2024)220",
    journal = "JHEP",
    volume = "02",
    pages = "220",
    year = "2024"
}

@article{nsubjettiness,
   title={Identifying boosted objects with {N}-subjettiness},
   volume={2011},
   ISSN={1029-8479},
   url={http://dx.doi.org/10.1007/JHEP03(2011)015},
   DOI={10.1007/jhep03(2011)015},
   number={3},
   journal={Journal of High Energy Physics},
   publisher={Springer Science and Business Media LLC},
   author={Thaler, Jesse and Van Tilburg, Ken},
   year={2011},
   month=mar }

@article{Mikuni:2025tar,
    author = "Mikuni, Vinicius and Nachman, Benjamin",
    title = "{Method to simultaneously facilitate all jet physics tasks}",
    eprint = "2502.14652",
    archivePrefix = "arXiv",
    primaryClass = "hep-ph",
    doi = "10.1103/PhysRevD.111.054015",
    journal = "Phys. Rev. D",
    volume = "111",
    number = "5",
    pages = "054015",
    year = "2025"
}

@article{Aad_2020,
  title         = {Search for heavy resonances decaying into a photon and a hadronically decaying {Higgs} boson in pp collisions at $\sqrt{s}=13$ {TeV} with the {ATLAS} detector},
  volume        = {125},
  issn          = {1079-7114},
  url           = {http://dx.doi.org/10.1103/PhysRevLett.125.251802},
  doi           = {10.1103/physrevlett.125.251802},
  number        = {25},
  journal       = {Physical Review Letters},
  publisher     = {American Physical Society (APS)},
  author        = "{ATLAS Collaboration}",
  year          = {2020},
  month         = dec
}

@article{Alwall:2011uj,
  author        = {Alwall, Johan and Herquet, Michel and Maltoni, Fabio and Mattelaer, Olivier and Stelzer, Tim},
  title         = {{MadGraph 5 : Going Beyond}},
  eprint        = {1106.0522},
  archiveprefix = {arXiv},
  primaryclass  = {hep-ph},
  reportnumber  = {FERMILAB-PUB-11-448-T},
  doi           = {10.1007/JHEP06(2011)128},
  journal       = {JHEP},
  volume        = {06},
  pages         = {128},
  year          = {2011}
}

@article{ATLAS_RPC_STOP_2020,
  title         = {Search for top squarks in events with a {Higgs} or {Z} boson using 139 fb$^{-1}$ of pp collision data at $\sqrt{s}=13$ {TeV} with the {ATLAS} detector},
  volume        = {80},
  issn          = {1434-6052},
  url           = {http://dx.doi.org/10.1140/epjc/s10052-020-08469-8},
  doi           = {10.1140/epjc/s10052-020-08469-8},
  number        = {11},
  journal       = {The European Physical Journal C},
  publisher     = {Springer Science and Business Media LLC},
  author        = "{ATLAS Collaboration}",
  year          = {2020},
  month         = nov
}

@article{xsh_2024,
  title         = {{Search for a resonance decaying into a scalar particle and a Higgs boson in the final state with two bottom quarks and two photons in proton-proton collisions at $\sqrt{s}$ = 13 TeV with the ATLAS detector}},
  volume        = {2024},
  issn          = {1029-8479},
  url           = {http://dx.doi.org/10.1007/JHEP11(2024)047},
  doi           = {10.1007/jhep11(2024)047},
  number        = {11},
  journal       = {Journal of High Energy Physics},
  publisher     = {Springer Science and Business Media LLC},
  author        = "{ATLAS Collaboration}",
  year          = {2024},
  month         = nov
}

@article{CMS_FCNC_2018,
  title         = {Search for the flavor-changing neutral current interactions of the top quark and the {Higgs} boson which decays into a pair of b quarks at $\sqrt{s}=13$ {TeV}},
  volume        = {2018},
  issn          = {1029-8479},
  url           = {http://dx.doi.org/10.1007/JHEP06(2018)102},
  doi           = {10.1007/jhep06(2018)102},
  number        = {6},
  journal       = {Journal of High Energy Physics},
  publisher     = {Springer Science and Business Media LLC},
  author        = "{CMS Collaboration}",
  year          = {2018},
  month         = jun
}

@article{collaboration_search_2020,
  title        = {Search for direct production of electroweakinos in final states with missing transverse momentum and a {Higgs} boson decaying into photons in $pp$ collisions at $\sqrt{s}=13$ {TeV} with the {ATLAS} detector},
  volume       = {2020},
  issn         = {1029-8479},
  url          = {http://arxiv.org/abs/2004.10894},
  doi          = {10.1007/JHEP10(2020)005},
  pages        = {5},
  number       = {10},
  journal = {Journal of High Energy Physics},
  journaltitle = {Journal of High Energy Physics},
  shortjournal = {J. High Energ. Phys.},
  author       = "{ATLAS Collaboration}",
  urldate      = {2025-08-17},
  year         = {2020},
  archivePrefix = "arXiv",
  eprint       = {2004.10894},
  primaryClass = "hep-ex"
}

@article{contrastive_embedding_bkgonly,
  title     = {Anomaly-preserving contrastive neural embeddings for end-to-end model-independent searches at the {LHC}},
  volume    = {112},
  issn      = {2470-0029},
  url       = {http://dx.doi.org/10.1103/5n77-ynsp},
  doi       = {10.1103/5n77-ynsp},
  number    = {7},
  journal   = {Physical Review D},
  publisher = {American Physical Society (APS)},
  author    = {Metzger, Kyle and Xu, Lana and Sodini, Mia and Årrestad, Thea K. and Govorkova, Katya and Grosso, Gaia and Harris, Philip},
  year      = {2025},
  month     = oct
}

@misc{finke_back_2023,
  title      = {Back To The Roots: Tree-Based Algorithms for Weakly Supervised Anomaly Detection},
  shorttitle = {Back To The Roots},
  number     = {{arXiv}:2309.13111},
  publisher  = {{arXiv}},
  author     = {Finke, Thorben and Hein, Marie and Kasieczka, Gregor and Krämer, Michael and Mück, Alexander and Prangchaikul, Parada and Quadfasel, Tobias and Shih, David and Sommerhalder, Manuel},
  year       = {2023},
  eprinttype = {arxiv},
  eprint     = {2309.13111},
  archivePrefix={arXiv},
  primaryClass = {hep-ph},
}

@misc{haxadv1,
  author        = {Cheng, Chi Lung and Demers, Sarah and Diefenbacher, Sascha and Li, Runze and Nachman, Benjamin and Noll, Dennis},
  title         = {{Weakly Supervised Anomaly Detection in Events with a Higgs Boson and Exotic Physics}},
  eprint        = {2508.13566},
  archiveprefix = {arXiv},
  primaryclass  = {hep-ex},
  month         = {8},
  year          = {2025}
}

@article{Monteux_2016,
  title     = {New signatures and limits on {R}-parity violation from resonant squark production},
  volume    = {2016},
  issn      = {1029-8479},
  url       = {http://dx.doi.org/10.1007/JHEP03(2016)216},
  doi       = {10.1007/jhep03(2016)216},
  number    = {3},
  journal   = {Journal of High Energy Physics},
  publisher = {Springer Science and Business Media LLC},
  author    = {Monteux, Angelo},
  year      = {2016},
  month     = mar
}

@misc{qu2024particletransformerjettagging,
  title         = {{Particle Transformer for Jet Tagging}},
  author        = {Huilin Qu and Congqiao Li and Sitian Qian},
  year          = {2024},
  eprint        = {2202.03772},
  archiveprefix = {arXiv},
  primaryclass  = {hep-ph},
}

@misc{simclr,
  title         = {{A Simple Framework for Contrastive Learning of Visual Representations}},
  author        = {Ting Chen and Simon Kornblith and Mohammad Norouzi and Geoffrey Hinton},
  year          = {2020},
  eprint        = {2002.05709},
  archiveprefix = {arXiv},
  primaryclass  = {cs.LG},
}

@misc{supervisedcontrastivelearning,
  title         = {{Supervised Contrastive Learning}},
  author        = {Prannay Khosla and Piotr Teterwak and Chen Wang and Aaron Sarna and Yonglong Tian and Phillip Isola and Aaron Maschinot and Ce Liu and Dilip Krishnan},
  year          = {2021},
  eprint        = {2004.11362},
  archiveprefix = {arXiv},
  primaryclass  = {cs.LG},
}

@article{SUSY-2018-31,
  title         = {{Search for bottom-squark pair production with the ATLAS detector in final states containing Higgs bosons, b-jets and missing transverse momentum}},
  volume        = {2019},
  issn          = {1029-8479},
  url           = {http://dx.doi.org/10.1007/JHEP12(2019)060},
  doi           = {10.1007/jhep12(2019)060},
  number        = {12},
  journal       = {Journal of High Energy Physics},
  publisher     = {Springer Science and Business Media LLC},
  author        = "{ATLAS Collaboration}",
  year          = {2019},
  month         = dec
}

@misc{vae,
  title         = {{Auto-Encoding Variational Bayes}},
  author        = {Diederik P Kingma and Max Welling},
  year          = {2022},
  eprint        = {1312.6114},
  archiveprefix = {arXiv},
  primaryclass  = {stat.ML},
  url           = {https://arxiv.org/abs/1312.6114}
}

@misc{buhmann2024phasespaceresonantanomaly,
      title={Full Phase Space Resonant Anomaly Detection}, 
      author={Erik Buhmann and Cedric Ewen and Gregor Kasieczka and Vinicius Mikuni and Benjamin Nachman and David Shih},
      year={2024},
      eprint={2310.06897},
      archivePrefix={arXiv},
      primaryClass={hep-ph},
}

@misc{grinsztajn2022treebasedmodelsoutperformdeep,
      title={Why do tree-based models still outperform deep learning on tabular data?}, 
      author={Léo Grinsztajn and Edouard Oyallon and Gaël Varoquaux},
      year={2022},
      eprint={2207.08815},
      archivePrefix={arXiv},
      primaryClass={cs.LG},
}

@article{PhysRevD.111.L051504,
  title = {Solving key challenges in collider physics with foundation models},
  author = {Mikuni, Vinicius and Nachman, Benjamin},
  journal = {Phys. Rev. D},
  volume = {111},
  issue = {5},
  pages = {L051504},
  numpages = {6},
  year = {2025},
  month = {Mar},
  publisher = {American Physical Society},
  doi = {10.1103/PhysRevD.111.L051504},
  url = {https://link.aps.org/doi/10.1103/PhysRevD.111.L051504}
}

@article{Dillon:2022tmm,
    author = "Dillon, Barry M. and Mastandrea, Radha and Nachman, Benjamin",
    title = "{Self-supervised anomaly detection for new physics}",
    eprint = "2205.10380",
    archivePrefix = "arXiv",
    primaryClass = "hep-ph",
    doi = "10.1103/PhysRevD.106.056005",
    journal = "Phys. Rev. D",
    volume = "106",
    number = "5",
    pages = "056005",
    year = "2022"
}

@article{JMLR:v9:vandermaaten08a,
  author  = {Laurens van der Maaten and Geoffrey Hinton},
  title   = {{Visualizing} {Data} using t-{SNE}},
  journal = {Journal of Machine Learning Research},
  year    = {2008},
  volume  = {9},
  number  = {86},
  pages   = {2579--2605},
  url     = {http://jmlr.org/papers/v9/vandermaaten08a.html}
}

@misc{cmscollaboration2025machinelearningtechniquesmodelindependentsearches,
      title={Machine-learning techniques for model-independent searches in dijet final states}, 
      author="{CMS Collaboration}",
      year={2025},
      eprint={2512.20395},
      archivePrefix={arXiv},
      primaryClass={hep-ex},
}

@misc{bhimji2025omnilearnedfoundationmodelframework,
      title={OmniLearned: A Foundation Model Framework for All Tasks Involving Jet Physics}, 
      author={Wahid Bhimji and Chris Harris and Vinicius Mikuni and Benjamin Nachman},
      year={2025},
      eprint={2510.24066},
      archivePrefix={arXiv},
      primaryClass={hep-ph},
}

\appendix

\section{Latent Space Information Preservation}
\label{app:iad_comparisons}

To quantify how much information the contrastive encoder retains, we compare the multi-class classification performance achieved with different feature representations.
Specifically, we train classifiers on the latent-space features produced by the encoder and on sets of raw physical features, and evaluate their performance on the same classification task.

\begin{figure*}[htbp]
  \centering
  \subfloat[Contrastive embedding BDT]{\includegraphics[width=0.48\textwidth]{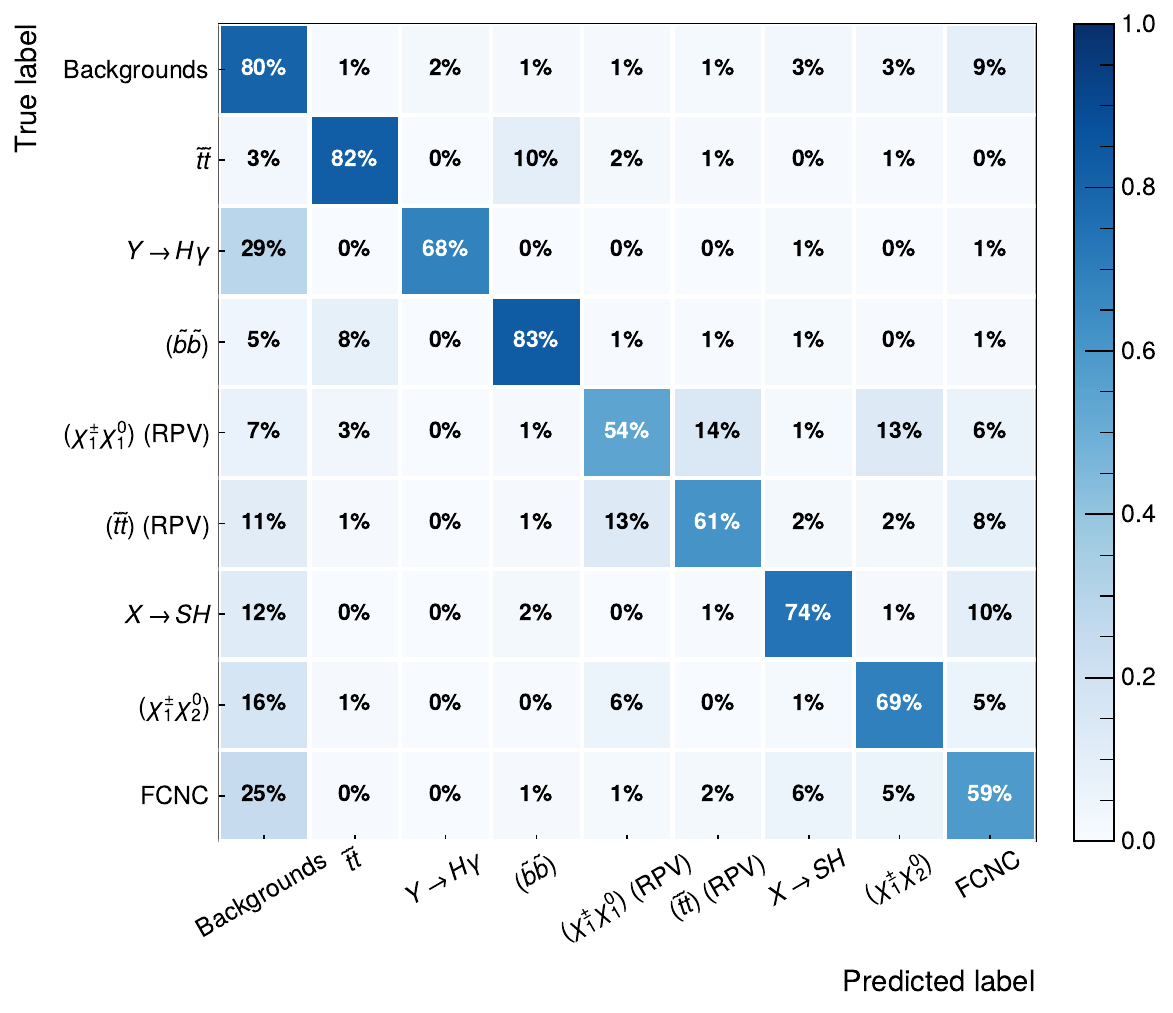}\label{fig:cm_encoded}}
  \hfill
  \subfloat[HAXADv1 BDT]{\includegraphics[width=0.48\textwidth]{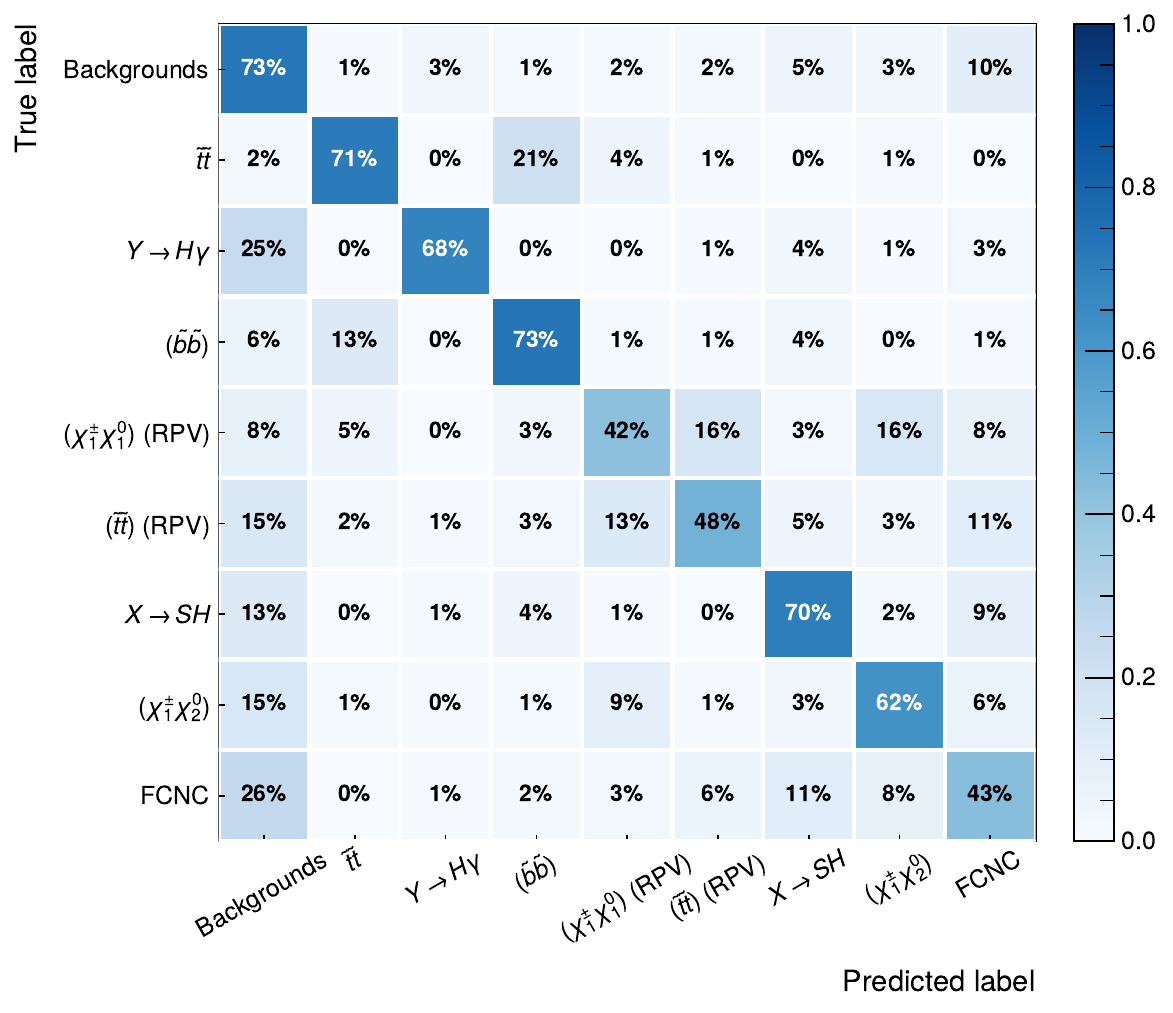}\label{fig:cm_haxad}}\\
  \subfloat[Supervised classifier]{\includegraphics[width=0.48\textwidth]{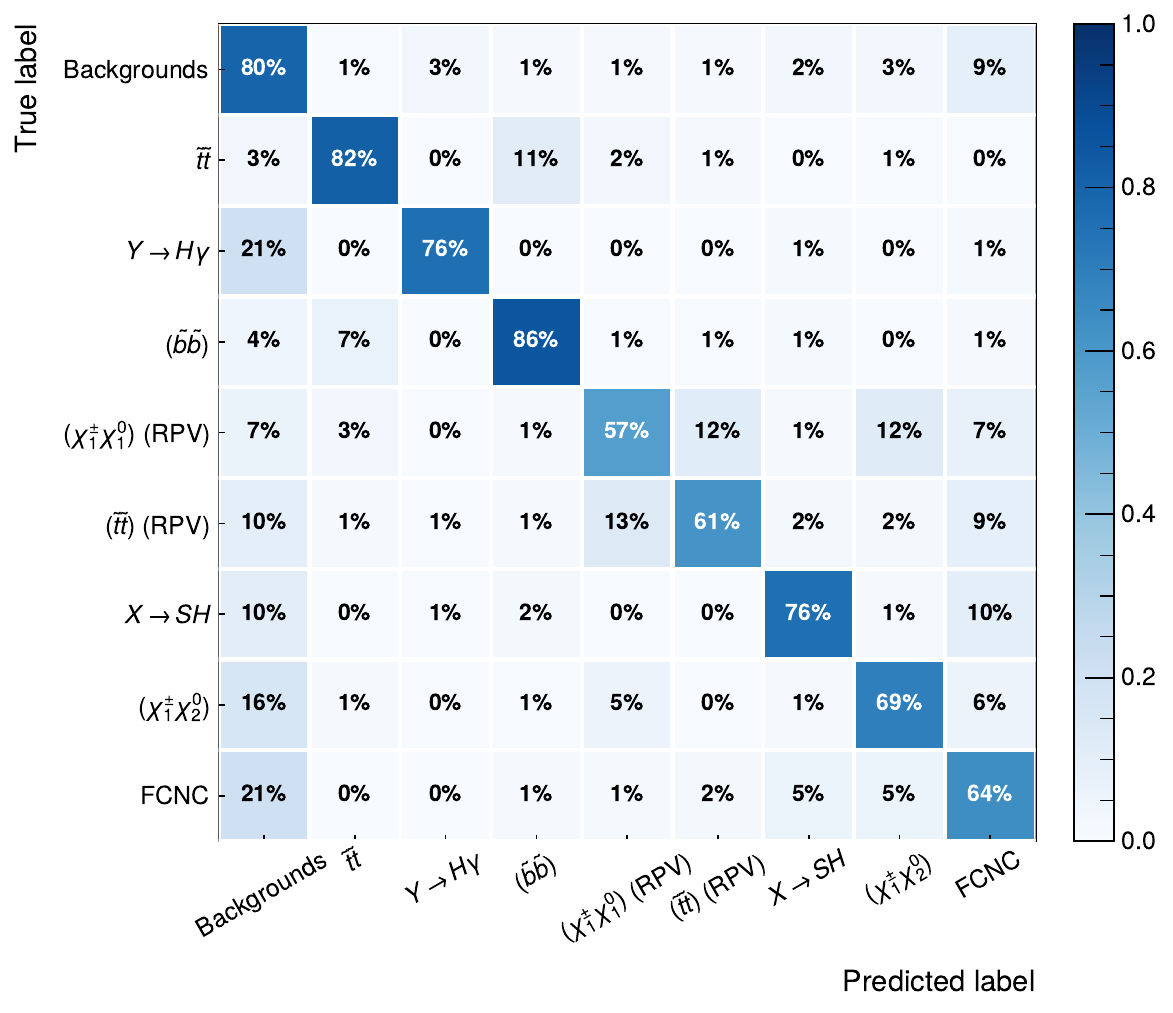}\label{fig:cm_supervised}}
  \caption{Confusion matrices comparing the classification performance of \protect\subref{fig:cm_encoded} a BDT trained on the contrastive embedding latent space features, \protect\subref{fig:cm_haxad} a BDT trained on nine raw physical features used in Ref. \cite{haxadv1}, and \protect\subref{fig:cm_supervised} a fully supervised particle transformer classifier.}
  \label{fig:iad_comparisons}
\end{figure*}

In this multi-class classification task, the classifiers are trained to distinguish among nine physics processes, including the \gls{SM} background processes as background classes and the eight \gls{BSM} processes listed in Table~\ref{tab:signals}.
The performance is shown in \cref{fig:iad_comparisons} in the form of confusion matrices, where three classifiers are compared.
The first is a BDT trained on the six-dimensional latent features from the contrastive encoder in the \gls{ID} setup (\cref{fig:cm_encoded}).
The second is a BDT trained on the nine raw physical features employed in Ref.~\cite{haxadv1}, corresponding to the best-case performance achievable with that feature set (\cref{fig:cm_haxad}).
The third uses the same particle transformer architecture as the contrastive encoder, but is trained in a fully supervised manner with the output and projection heads replaced by an \gls{MLP} with a softmax activation function (\cref{fig:cm_supervised}).
The training schedule is kept the same as in Section~\ref{sec:embedding}, except that the loss function is replaced with the cross-entropy loss.
This setup represents the upper bound on the classification performance achievable with our encoder architecture.

Relative to the fully supervised particle transformer, the BDT trained with latent space features exhibits only a small reduction of about 5\% in the diagonal entries of the confusion matrix, indicating limited information loss from the contrastive encoding. 
By contrast, the degradation increases to about 10\% for the BDT based on the nine raw physical features used in Ref.~\cite{haxadv1}. 
In particular, the background classification accuracy is nearly unchanged for the latent-space classifier, while it decreases by roughly 7\% for the BDT trained on the raw physical features. 
These results suggest that the contrastive encoder retains most of the information relevant for distinguishing \gls{BSM} signals from \gls{SM} backgrounds, which is one key requirement for a high anomaly detection performance.

\end{document}